\documentclass[twocolumn,journal]{IEEEtran}
\usepackage[noadjust]{cite}

\ifCLASSINFOpdf
   \usepackage[pdftex]{graphicx}
\else
   \usepackage[dvips]{graphicx}
\fi
\usepackage[colorlinks,
            linkcolor=black,
            urlcolor=red,
            anchorcolor=black,
            citecolor=black
            ]{hyperref}
\usepackage{tabularx}
\usepackage{supertabular}
\usepackage{amsmath,amssymb,amsfonts}
\usepackage{mathrsfs}
\usepackage{multirow}
\usepackage{algorithm}
\usepackage{algpseudocode}
\usepackage{bm}
\usepackage{longtable}
\usepackage{xcolor}
\usepackage{lineno}
\usepackage{tikz}
\usepackage{circuitikz}
\usepackage{fixltx2e}
\usepackage{url}

\begin{document}
\title{Characterizing the Burst Error Correction Ability of Quantum Cyclic Codes}
%

\author{Jihao~Fan,~\IEEEmembership{Member,~IEEE}~and Min-Hsiu Hsieh,~\IEEEmembership{Senior Member,~IEEE}
        \thanks{This work was supported by the National Natural Science Foundation of China under Grants  62371240, and 61802175. This paper was presented in part at the  IEEE International Symposium on
Information Theory, Vail, USA, June 2018.  }
\thanks{J. Fan is with School of Cyber Science and Engineering, Nanjing University of Science and Technology, Nanjing 210094, China (e-mail: jihao.fan@outlook.com).  }
\thanks{M. H. Hsieh   is  with Hon Hai Quantum Computing Research Center,   Taipei 114, Taiwan (email: minhsiuh@gmail.com)}
 }

\maketitle

\begin{abstract}
Quantum burst error correction codes (QBECCs) are of great importance to deal with the memory effect  in quantum channels. As the most important family of QBECCs,  quantum cyclic codes (QCCs) play a vital role in the correction of burst errors. In this work, we characterize the burst error correction ability  of QCCs constructed from the Calderbank-Shor-Steane (CSS) and the Hermitian constructions. We  determine the burst error correction limit of  QCCs and quantum Reed-Solomon codes with algorithms in polynomial-time complexities.  As a result, lots of QBECCs saturating the quantum Reiger bound are obtained.  We  show that        quantum Reed-Solomon codes   have better burst error correction abilities than the previous results. At last, we give the quantum error-trapping decoder (QETD) of QCCs for decoding  burst errors.  The   decoder runs in linear time and    can decode both degenerate and nondegenerate   burst errors. What's more, the numerical results show that   QETD can decode much more degenerate burst errors than the nondegenerate ones.
\end{abstract}


%
\IEEEpeerreviewmaketitle

\section{Introduction}
\newcounter{conter1}
\setcounter{conter1}{0}
\newtheorem{theorem}[conter1]{Theorem}

\newtheorem{definitions}{Definition}
\newtheorem{theorems}{Theorem}
\newtheorem{lemmas}{Lemma}
\newtheorem{corollarys}{Corollary}
\newtheorem{examples}{Example}
\newtheorem{propositions}{Proposition}
%
%
%
%

\IEEEPARstart{N}{oise} induced by decoherence is a main obstacle in the    realization of quantum computers and quantum communications. One of the most usual and important methods to deal with such phenomenon  is  using     quantum error correction codes (QECCs) to correct the errors \cite{nielsen2010quantum}. In standard quantum coding theory, quantum errors are assumed to be discrete and independent with each other. However, in   practical quantum information processing systems, quantum errors   tend to be correlated in space and time \cite{caruso2014quantum,clemens2004quantum,jacobsen2014optimal,chiribella2011quantum,mcewen2022resolving}.
Thus quantum channels usually have a memory effect and bring in errors which are localized, e.g.,   bursts of errors.

Similar to classical coding theory \cite{lin2004error}, there are    quantum burst error correction codes (QBECCs) {\cite{jihao2017on, vatan1999spatially, kawabata2000quantum, tokiwa2005some}} for correcting   bursts of errors.
In \cite{vatan1999spatially}, the spatially correlated qubit errors are considered and the first  families of QBECCs  were constructed by using the Calderbank-Shor-Steane (CSS) construction \cite{calderbank1996good, steane1996error}. In \cite{tokiwa2005some}, QBECCs of short block length were found by using computer exhaustive search. In \cite{kawabata2000quantum,Trinca2022New},  quantum interleavers for QBECCs were proposed so that long QBECCs can be produced  from short ones. In \cite{qian2015constructions}, some QBECCs were obtained from quantum Reed-Solomon (RS) codes. In \cite{fan2018constructionISIT}, the framework of general QBECCs was presented and \emph{degenerate} QBECCs were first proposed. In \cite{trinca2023construction,trinca2024new}, quantum burst error correction was related to topological quantum error-correcting codes and new QBECCs were constructed by using interleaving techniques.  In \cite{con2025qbel}, a new class of quantum burst error-locating codes was proposed to locate and correct burst errors.

Compared to the development of standard QECCs \cite{panteleev2022asymptotically,leverrier2022quantum,dinur2023good} or entanglement-assisted QECCs \cite{7501532, Brun436, 6461941, 6671483, 5714249, PhysRevA.79.032340, PhysRevA.76.062313,fan2022entanglement,fan2016constructions}, the construction and investigation of QBECCs have received far less attention. Although   standard QECCs can also be used to  correct burst errors, they are not efficient enough and usually have a much smaller code rate than QBECCs for correcting burst errors of the same length. The current QBECCs are mainly obtained with interleaving or by using   computer exhaustive search. The interleaving technology highly relies  on short QBECCs and the parameters are restricted to some specific numbers. The time complexity of exhaustively  searching of general QBECCs are exponential to the code length \cite{lin2004error}.  In addition,  there is an interesting class of quantum codes, called  {degenerate} codes, that have no classical correspondences. Degenerate codes can potentially  store more quantum information or correct more quantum errors than nondegenerate ones \cite{nielsen2010quantum,fan2022partially,fan2021asymmetric,fan2022entanglement}.  However,   determining the burst error correction ability of degenerate QBECCs  is a quite difficult problem \cite{fan2018constructionISIT}. In particular,   QBECCs saturating the quantum Reiger bound is the  state-of-the-art   in coding theory \cite{lin2004error,fan2018constructionISIT}.
On the other hand,  almost all of  quantum burst error correction codes are   quantum cyclic codes (QCCs), which can be efficiently  realized by using the quantum shift register  \cite{grassl2000cyclic}. This phenomenon learns from classical coding theory, in which   cyclic codes are much more practical than non-cyclic codes \cite{lin2004error}.
How to determine the burst error correction limit of general QCCs is still unknown. Meanwhile, quantum RS codes are efficient to correct both random and burst errors. There exists   the lower bound for the burst error correction ability of    quantum RS codes \cite{qian2015constructions,lin2004error}. Whether the true burst error correction limit of     quantum RS codes can exceed the lower bound is unknown.

  In this paper we characterize the burst error correction ability of QCCs by generalizing the algorithms in \cite{matt1980determining} for   classical cyclic codes to QCCs.   We propose polynomial-time algorithms    to determine the burst error correction limit of general QCCs. Moreover, degenerate errors are particularly considered in  the algorithms. As a result, a lot of new QBECCs which can achieve the quantum Reiger bound are obtained by running the polynomial-time searching algorithms in Magma software  (V2.28-3) \cite{cannon2023handbook}. What's more, we also propose polynomial-time algorithms to determine the true burst error correction limit of quantum RS codes. Many quantum RS codes with burst error correction ability beating the lower bound in  \cite{qian2015constructions,lin2004error} are derived.   At last, we propose the quantum error-trapping decoder (QETD) for  correcting burst errors of QCCs. This decoding algorithm runs in  linear-time  and is  quantum maximum likelihood
decoding. In addition, the quantum error-trapping decoder can not only decode all the burst errors that are coset leaders but also can correct     degenerate errors which belong to the coset of coset leaders. Our numerical results show that QETD can decode much more degenerate burst errors than the nondegenerate ones. The Magma   codes for computing the burst error correction limit of QCCs and quantum RS codes, and evaluating the performance of QETD are put in \cite{githubjihao}.

The rest of the paper is organized as follows.  In Section \ref{Preliminaries}, we give some basic knowledge of QBECCs and QCCs.  Section \ref{polynomialtimealgorithmofQCCs} proposes  the polynomial-time algorithm for determining the burst error correction ability of   QCCs. The polynomial-time algorithm for  determining the burst error correction ability of   quantum RS codes is given in Section \ref{polynomialalgorithmburstquantumRScodes}. Section \ref{quantumerrortrappingsection} presents the quantum error-trapping decoder for QCCs. The conclusion and the discussion are given in Section \ref{conclusionanddiscussion}.


\section{Preliminaries}
\label{Preliminaries}
In this section, we present some basic definitions and  backgrounds of QECCs and develop the stabilizer formalism for  QBECCs. For simplicity and  practice, we mainly  consider the qubit system in this paper.

Denote  the complex Hilbert space by $\mathbb{C}^2$. A qubit  $|v\rangle \in \mathbb{C}^2$ can be written as
$
|v\rangle=\alpha|0\rangle+\beta|1\rangle,
$
where $\alpha$ and $\beta$ are complex numbers satisfying $|\alpha|^2+|\beta|^2=1$.  An $n$-qubit $|\psi\rangle$ is then a quantum state in the $n$-th tensor product   of $\mathbb{C}^2$, i.e., $|\psi\rangle \in \mathbb{C}^{2^n}\equiv \mathbb{C}^2\otimes\mathbb{C}^2\otimes\cdots\otimes\mathbb{C}^2$. The Pauli matrices
\begin{equation*}
\label{Pauli_Matrices}
I_2=  \left[
\begin{matrix}
1&0\\
0&1\\
\end{matrix}
\right],
X=  \left[
\begin{matrix}
0&1\\
1&0\\
\end{matrix}
\right],
Z=  \left[
\begin{matrix}
\setlength{\arraycolsep}{0.1pt}
1&0\\
0&-1\\
\end{matrix}
\right],
Y=  \left[
\begin{matrix}
\setlength{\arraycolsep}{0.1pt}
0&-i\\
i&0\\
\end{matrix}
\right]
\end{equation*}
form a basis of the linear operators on $\mathbb{C}^2$.
Let $q $  be a power of a prime $p$ and let $\mathbb{F}_p$ be  the prime field.  Let $m\geq1$ be an integer. Let $\mathbb{F}_q$ be the Galois field  with $q$ elements and let the   field $ \mathbb{F}_{q^m} $    be a field extension  of $\mathbb{F}_q$.      The trace operation from $\mathbb{F}_{q^m}$ to $\mathbb{F}_q$ is  defined as $\textrm{Tr}(\alpha)=\sum_{i=0}^{m-1}\alpha^{q^i}$.

According to the   discretized  model for  quantum errors (see \cite{nielsen2010quantum, 6778074}), we only need to consider a discrete set of quantum errors of $n$ qubits.
Further,   the bit-flip error  ($X$-error), the phase-flip error  ($Z$-error), and the combined bit-flip and phase-flip  error  ($Y$-error) are three basic   errors in quantum channels. Then
 the  error group is defined as follows
\begin{equation}
\mathcal{G}_n=\{i^\lambda w_1\otimes\cdots\otimes w_n|0\leq\lambda\leq 3, w_i\in{I_2,X,Y,Z}\}.
\end{equation}
Furthermore, it is sufficient to consider the quotient group $\mathcal{\overline{G}}_n=  \mathcal{G}_n/\{\pm1,\pm i\}$ of  $\mathcal{G}_n$ since the global phase  $i^\lambda$ in $\mathcal{G}_n$ is not needed. Let $\overline{e}= w_1\otimes w_2\otimes\cdots\otimes w_n\in\mathcal{\overline{G}}_n$ and let  $e=i^\lambda\overline{e}\in\mathcal{G}_n$. The  burst  length of $\bar{e}$ to be $\ell$  is denoted  by $\textrm{bl}(\overline{e})=\ell$, where the nonidentity matrices in  $\overline{e}$ are confined to  $\ell$ consecutive positions.

The idea of a QECC is to encode quantum information into a  subspace of some larger Hilbert space. An $[[n,k]]$ QECC $Q$ is defined to be the subspace of dimension $2^k$ in $\mathbb{C}^{2^n}$.   According to the error correction conditions of QECCs   in  \cite{knill1997theory},  the error correction condition of QECCs for correcting burst errors  is given as follows

\begin{propositions}
\label{proposition_burst}
The quantum code $Q$ can correct any quantum    burst error  of length $l$ or less if and only if
\begin{equation}
\label{burst-error-correction criterion}
\langle c_i|E^\dagger E'|c_j\rangle=a_{(E,E')}\delta_{ij}
\end{equation}
for all $\langle c_i|c_j\rangle = \delta_{ij}$ and for all $\textrm{bl}(E),\textrm{bl}(E')\leq \ell$, where $|c_i\rangle$ and $|c_j\rangle\in Q$, $E$ and $E'\in\mathcal{G}_n$, and $a_{(E,E')}$ is a constant     which depends only on $E$ and $E'$.
If $\langle c_i|E^\dagger E'|c_j\rangle=0$ for all $|c_i\rangle,|c_j\rangle\in Q$ and for all $\textrm{bl}(E),\textrm{bl}(E')\leq \ell$, where $E\neq E' \in\mathcal{G}_n$, then $Q$ is a  nondegenerate QBECC.
\end{propositions}

Similar with the group theoretical framework for standard  QECCs in  \cite{calderbank1997quantum,calderbank1998quantum},  the stabilizer formalism for QBECCs was given in \cite{fan2018constructionISIT}.
Furthermore, the  CSS and the Hermitian constructions \cite{calderbank1996good,steane1996error,calderbank1998quantum} provide a  more direct way to construct QECCs from classical linear or additive codes than
Proposition
\ref{proposition_burst}.
 \begin{lemmas}{\cite{fan2018constructionISIT}}
\label{CSS QBECCs}
Let $C_1=[n,k_1]$ and $C_2=[n,k_2]$ be
two binary linear codes satisfying  $C_2^\bot\subseteq C_1$. Suppose that     $\ell$ is the largest integer such that   $e_1+e_2\notin  (C_1 \backslash  C_2^\bot) \cup (C_2 \backslash  C_1^\bot)$    for  arbitrary two binary vectors $e_1\neq e_2 $ with $0\leq \textrm{bl}(e_1) $, $\textrm{bl}(e_2)\leq \ell$.
   There exists a \[Q=[[n,k_1+k_2-n]]\] binary QBECC   which can correct arbitrary    quantum error  of burst length $\ell$ or less. For all the $0\leq \textrm{bl}(e_1) $, $\textrm{bl}(e_2)\leq \ell$, if    $e_1+e_2\notin  (C_1   \cup  C_2) \backslash  \{0\}$, then $Q$ is a nondegenerate code, otherwise it is degenerate.
\end{lemmas}

\begin{lemmas}{\cite{fan2018constructionISIT}}
\label{Hermitian QBECCs}
Let $C=[n,k]_4$ be  an additive code over $\mathbb{F}_4$ and suppose that $C^{\bot_H}\subseteq C $, where $C^{\bot_H}$ is the Hermitian dual of $C$. Suppose that     $\ell$ is the largest integer such that   $e_1+e_2\notin C  \backslash  C^{\bot_H}$ for  arbitrary two vectors $e_1\neq e_2\in\mathbb{F}_4^{n}$ with $0\leq \textrm{bl}(e_1) $, $\textrm{bl}(e_2)\leq \ell$. There exists a
\[Q=[[n,2k-n]]\] binary QBECC   which can correct any quantum burst error  of length $\ell$ or less. For all the $0\leq \textrm{bl}(e_1) $, $\textrm{bl}(e_2)\leq \ell$, if  $e_1+e_2\notin C  \backslash  \{0\}$, then $Q$ is a nondegenerate  QBECC, otherwise it is degenerate.
\end{lemmas}

For a classical burst error correction code  $C=[n,k]$  which can correct  any burst errors of length $\leq l$,   there exists an important upper bound called the Reiger bound: $n-k\geq 2\ell$ that constrains the burst error correction ability of $C$ (see \cite{lin2004error}).   Let  $Q=[[n,k]]$ be a QECC which can correct any quantum random error of length up to $ t$, then there exists the quantum Singleton bound  $n-k\geq 4t$  which  is an upper bound for the quantum random error correction ability of code $Q$ (see \cite{nielsen2010quantum,calderbank1998quantum}).
In the following, we derive the quantum Reiger bound (QRB) which is   an upper bound for the quantum burst error correction ability of code $Q$.

\begin{theorems} [Quantum Reiger Bound]
\label{quantumReigerbound}
If an $[[n,k]]$ QBECC $Q$ can correct quantum burst errors of length $\ell$, then there is
\begin{equation}
n-k\geq 4\ell.
\end{equation}
\end{theorems}



\begin{IEEEproof}
The proof is given in Appendix \ref{appendixA}.
\end{IEEEproof}

For an $[[n,k]]$ QBECC $Q$ which can correct any burst error of length up to $\ell$, we denote by $Q=[[n,k;\ell]]$. If $Q$ can saturate the quantum Reiger bound, i.e., $n-k-4\ell=0$, then we say $Q$ is \emph{optimal}. If $n-k-4\ell =1$ or $n-k-4\ell =2$, then we say $Q$ is \emph{nearly optimal}. In the following section, we will  focus on   QCCs which are optimal or nearly optimal. In this paper, we use a subscript $q$ in the parameters of both classical  and quantum codes to represent the finite field. If $q=2$, then we omit the subscript in the parameters of both classical  and quantum codes  provided   there do not exist ambiguities.

\section{The Burst Error Correction Ability of Quantum Cyclic Codes}
\label{polynomialtimealgorithmofQCCs}

In coding theory, it is one of the central questions to determine the error correction ability of a code. However, this problem is usually  difficult from the perspective of  computational complexity. For example, it is   NP-hard  to compute a code's   minimum distance   \cite{vardy1997intractability}, which characterizes the random error correction ability of a code. While for determining  the    burst error correction ability of a code, the exhaustive searching is needed and the time is generally exponential with the code length \cite{lin2004error}. In the regime of quantum codes, the error correction issues are also difficult and seem to be even harder due to   error degeneracy   \cite{hsieh2011np,iyer2015hardness}. Nevertheless, for some specific codes, e.g., the quantum  cyclic codes,  we can determine  the burst error correction ability in polynomial-time complexity.
  We generalize the algorithms  in  Ref.~\cite{matt1980determining} to   the quantum regime, in which    error degeneracy is particularly considered.

We  construct quantum cyclic codes from classical cyclic codes satisfying the dual containing relationship in Lemma \ref{CSS QBECCs} and Lemma \ref{Hermitian QBECCs}. It should be noted that  we mainly give the detailed process of determining  the burst error correction ability of  Hermitian-type QCCs. For other QCCs, e.g., the CSS-type QCCs, the process is simlilar to that of the Hermitian-type QCCs.  
Let ${C}=[n,k]_{q^2}$ be   a classical cyclic code over $\mathbb{F}_{q^2}$. Denote  $r=n-k$ by the number of check symbols. Denote   $\mathbf{g}(x)=g_0+g_1x+\cdots+ g_{r-1}x^{r-1}+g_rx^r$ and $\mathbf{h}(x)=h_0+h_1x+\cdots+ h_{k-1}x^{k-1}+ h_kx^k$ by the generator and the parity-check polynomials, respectively. The generator matrix of ${C}$ is given by
\begin{equation}
\label{generatormatrix1}
\mathbf{G} =
\left(
\begin{array}{ccccccccc}
 g_r&g_{r-1}&\cdots &g_1&g_0&0&\cdots&0 \\
0&g_r&\cdots &g_2&g_1&g_0&\cdots&0 \\
 \vdots&\vdots&\ddots&\ddots&\ddots&\ddots&\ddots&\vdots \\
0&0&\cdots&g_r&\cdots &g_{i}&\cdots&g_0
 \end{array}
\right),
\end{equation}
where $1\leq i\leq r-1$.
The parity-check matrix of ${C}$ is given by
\begin{equation}
\label{paritycheckmatrix1}
\mathbf{H} =
\left(
\begin{array}{ccccccccc}
 h_0&h_1&\cdots &h_{k-1}&h_k&0&\cdots&0 \\
0&h_0&\cdots &h_{k-2}&h_{k-1}&h_k&\cdots&0 \\
 \vdots&\vdots&\ddots&\ddots&\ddots&\ddots&\ddots&\vdots \\
0&0&\cdots&h_0&\cdots &h_{j}&\cdots&h_k
 \end{array}
\right),
\end{equation}
where $1\leq j\leq k-1$.
   Let  $\mathcal{M}^{(t)} $ be the $(r-t)\times(n-t)$  matrix formed by deleting
the last $t$ rows and   last $t$ columns of the parity-check
matrix $\mathbf{H}$ in (\ref{paritycheckmatrix1}), where $1\leq t\leq (n-k)/2$.
Let $A=(a_{ij})$ be a matrix with elements over  $\mathbb{F}_{q^2}$, where $1\leq i\leq m$ and $1\leq j\leq n$. The conjugate transpose of $A$ is given by $A^\dagger=(b_{uv}^q)$, where $b_{uv}^q =a_{vu}^q$, $1\leq u\leq n$ and $1\leq v\leq m$. Then the conjugate transpose of $\mathbf{G}$ and $\mathbf{H}$ is denoted by $\mathbf{G}^\dagger$ and $\mathbf{H}^\dagger$, respectively. Firstly, we need the following three lemmas for   determining  the burst error correction ability of QCCs.

\begin{lemmas}
{\cite{calderbank1998quantum,ketkar2006nonbinary}}
Let $\textrm{gcd}(n,q^2)=1$ and let ${C}=[n,k]_{q^2}$ be a classical cyclic code. If the parity-check matrix $\mathbf{H} $ of ${C}$ satisfying $\mathbf{H}\mathbf{H}^{\dagger}=0$, then ${C}^{\bot_H}\subseteq {C}$ and there exists a ${Q}=[[n,2k-n]]_q$ QCC.
\end{lemmas}
\begin{lemmas}[{\cite[Theorem 1]{matt1980determining}}]
\label{classicalburstcyclic}
Let ${C} $ be a classical cyclic code over $\mathbb{F}_{q^2}$ with the parity-check matrix given in (\ref{paritycheckmatrix1}). Let $L$ be the largest integer $1\leq b\leq r$ such that every set of $b$   consecutive columns of the matrix $\mathcal{M}^{(b)} $ is linearly independent. Then ${C}$ can correct any burst error of length $L$ or less.
\end{lemmas}
\begin{lemmas}[\cite{lay2011linear}]
\label{diagonalmatrixformlemma}
Let $M$ be a matrix  of size $m\times n$. Then $M$ can always be transformed to the following form
\begin{equation}
\label{diagonalmatrixform}
\widehat{M} =
\left(
\begin{array}{cc}
\mathbf{I}&X  \\
\mathbf{0}&\mathbf{0}
 \end{array}
\right)
\end{equation}
through  elementary row operations and a permutation of columns (if necessary), where $\mathbf{I}$ is an identity matrix of dimension
 $r$ equal to the rank of $M$, $X$ is a matrix with
$r$ rows and
$n-r$ columns, and the two $\mathbf{0}$'s are zero matrices.
\end{lemmas}

For a nonzero matrix $M$ of size $m\times n$ and a set $A=[i,j]$ with $1\leq i\leq j\leq n$,
 we define the set  of $j-i+1$ consecutive columns  of $M$ indexed by $A $ as $M_A$. Denote the set of columns of $M$ by $\{m_i|1\leq i\leq n\}$, i.e., $M=(m_i)_{1\leq i\leq n}$.  We say that a set $\mathscr{S}$ of columns of $M$ is \emph{maximally linearly independent} (MLI) if including any other column in $M$ would make it linearly dependent. If all the columns of $M$ is  linearly independent, then $\mathscr{S}=\{m_i|1\leq i\leq n\}$. It is known that the rank of $M$ is equal to the number of elements in $\mathscr{S}$.

 Let $1\leq \ell\leq r$ and let $A_{\imath}=[\imath,\imath+\ell-1]$, where $1\leq \imath\leq n-2\ell+1$.   Denote   a set of MLI columns in  $\mathcal{M}^{(\ell)}_{A_{\imath}}$ by $\mathscr{D}_H=\{\alpha_1,\cdots,\alpha_{u}\}$, where $\alpha_u\in\mathcal{M}^{(\ell)}_{A_{\imath}} $ and $1\leq u\leq \ell$ is the rank of $\mathcal{M}^{(\ell)}_{A_{\imath}}$. We define $\widehat{\mathscr{D}}_H= \{\beta_1,\cdots,\beta_v\}$  as the complementary set of $\mathscr{D}_H$, where $v=\ell-u$. If $\mathcal{M}^{(\ell)}_{A_{\imath}}$  is of full rank, then $u=\ell$ and $\widehat{\mathscr{D}}_H$ is empty. Otherwise,  each element in $\widehat{\mathscr{D}}_H$ can be represented by a linear combination of vectors in $\mathscr{D}_H$, i.e., $\beta_i=a_{i1}\alpha_1+\cdots+a_{iu}\alpha_u$ for $\beta_i\in\mathscr{D}_H$ and $1\leq i\leq v$. Denote by  a burst error  $e_i$ ($1\leq i\leq v$) whose nonzero components fall in   the positions of $\{\alpha_1,\cdots,\alpha_u,\beta_i\}$ in $\mathbf{H}$.
  Then there exists a burst error $f_i$ whose nonzero components falling entirely   in the last $\ell$  positions such that $ e_i+f_i\in {C}$ for each $1\leq i\leq v$ according to Lemma \ref{classicalburstcyclic}.  We define a set of $e_i$ and $f_i$ as
 \begin{equation}
  \boxplus_{\mathcal{M}^{(\ell)}_{A_{\imath}}}=\{(e_1,f_1),\cdots,(e_v,f_v)\}.
 \end{equation}

\newcommand{\tabincell}[2]{\begin{tabular}{@{}#1@{}}#2\end{tabular}}

 \begin{table*}
\caption{Computer searching for optimal and nearly optimal nondegenerate  QCCs of length $n<100$.  }
\label{Computer Searching for QBECCs with Alg1}
\centering
\begin{tabular}{|l|l|l|l| }
\hline
 $\Delta$&  $[[n,k]]$&$\mathcal{L}$&Generator Polynomials
 \\
 \hline
 \multirow{29}{*}{$0$} &   $[[13,1]]$&   $3$& $g=(\textbf{1}^{6}\textbf{2}^{5}\textbf{3}^{3}\textbf{2}^1\textbf{1}^0)$   \\
  \cline{2-4}
  &  $[[15,3]]$ &   $3$& $g=(\textbf{1}^{6}\textbf{2}^{3}\textbf{1}^0)$   \\
  \cline{2-4}
  &  $[[17,1]]$ &   $3$& $g=(\textbf{1}^{8}\textbf{3}^{7}\textbf{1}^{6}\textbf{1}^{5}\textbf{2}^{4}\textbf{1}^{3}\textbf{1}^{2}\textbf{3}^{1}\textbf{1}^0)$   \\
\cline{2-4}
  &  $[[25,5]]$ &   $5$& $g=(\textbf{1}^{10}\textbf{2}^{5}\textbf{1}^0)$   \\
  \cline{2-4}
  &  $[[35,7]]$ &   $7$& $g=(\textbf{1}^{14}\textbf{3}^{7}\textbf{1}^0)$   \\
  \cline{2-4}
  &  $[[39,3]]$ &   $9$& $g=(\textbf{1}^{18}\textbf{3}^{15}\textbf{2}^{9}
  \textbf{3}^{3}\textbf{1}^0)$   \\
   \cline{2-4}
  &  $[[45,9]]$ &   $9$& $g=(\textbf{1}^{18}\textbf{2}^{9} \textbf{1}^0)$   \\
    \cline{2-4}
  &  $[[51,19]]$ &   $8$& $g=(\textbf{1}^{16}\textbf{2}^{14}\textbf{2}^{12}\textbf{1}^{8}\textbf{2}^{6} \textbf{2}^{5}\textbf{3}^{4}\textbf{3}^{3}\textbf{3}^{2}\textbf{3}^{1}\textbf{1}^0)$   \\
      \cline{2-4}
  &  $[[63,15]]$ &   $12$& $g=(\textbf{1}^{24}\textbf{2}^{21}\textbf{2}^{20}\textbf{2}^{14}\textbf{2}^{12}\textbf{1}^{8}\textbf{2}^{6} \textbf{2}^{5}\textbf{3}^{4}\textbf{3}^{3}\textbf{3}^{2}\textbf{3}^{1}\textbf{1}^0)$   \\
    \cline{2-4}
  &  $[[63,27]]$ &   $9$& $g=(\textbf{1}^{18}\textbf{3}^{17}\textbf{2}^{16}\textbf{1}^{15}\textbf{2}^{14}\textbf{2}^{13}\textbf{3}^{12}\textbf{1}^{11}\textbf{2}^{10}\textbf{3}^{9}\textbf{1}^{6} \textbf{1}^{5}\textbf{1}^{3}\textbf{3}^{1}\textbf{3}^{0}$   \\
     \cline{2-4}
  &  $[[65,5]]$ &   $15$& $g=( \textbf{1}^{30}\textbf{3}^{25}\textbf{2}^{15}\textbf{3}^{5} \textbf{1}^0)$
  \\
     \cline{2-4}
  &  $[[65,13]]$ &   $13$& $g=( \textbf{1}^{26}\textbf{3}^{13}\textbf{1}^0)$
  \\
    \cline{2-4}
  &  $[[65,29]]$ &   $9$& $g=(\textbf{1}^{18}\textbf{3}^{17}\textbf{1}^{15}\textbf{3}^{13}\textbf{3}^{12}\textbf{2}^{11}\textbf{3}^{9}\textbf{2}^{7}\textbf{3}^{6}\textbf{3}^{5}\textbf{1}^{3}
 \textbf{3}^{1} \textbf{1}^0)$
  \\
 \cline{2-4}
  &  $[[75,15]]$ &   $15$& $g=( \textbf{1}^{30}\textbf{2}^{15}\textbf{1}^0)$
  \\
     \cline{2-4}
  &  $[[85,9]]$ &   $19$& $g=(\textbf{1}^{38}\textbf{1}^{36}\textbf{3}^{35}\textbf{3}^{34}\textbf{1}^{33}\textbf{1}^{31}\textbf{1}^{29}\textbf{2}^{28}\textbf{3}^{27}\textbf{3}^{26}\textbf{3}^{24}\textbf{1}^{23}\textbf{3}^{22}\textbf{3}^{21}\textbf{1}^{20}\textbf{3}^{19}\textbf{1}^{18}\textbf{1}^{17}\textbf{1}^{16}\textbf{3}^{14}\textbf{3}^{13}\textbf{2}^{10}\textbf{3}^{8}\textbf{1}^{6}\textbf{2}^{5}\textbf{1}^{4}\textbf{2}^{3} \textbf{2}^{2} \textbf{1}^{1}\textbf{1}^{0})$   \\
   \cline{2-4}
  &  $[[85,17]]$ &   $17$& $g=(\textbf{1}^{34}\textbf{1}^{33}\textbf{2}^{32}
  \textbf{2}^{31}\textbf{3}^{30}\textbf{3}^{29}\textbf{1}^{28}
  \textbf{3}^{26}\textbf{2}^{25}\textbf{3}^{24}\textbf{3}^{23}
  \textbf{2}^{22}\textbf{1}^{20}\textbf{1}^{19}\textbf{3}^{18}
  \textbf{1}^{17}\textbf{2}^{13}\textbf{2}^{12}\textbf{1}^{9}
  \textbf{3}^{8}\textbf{2}^{7}\textbf{2}^{6}\textbf{1}^{4}
  \textbf{2}^{3}\textbf{2}^{2}\textbf{3}^{1}\textbf{1}^{0})$   \\
  \cline{2-4}
  &  $[[85,25]]$ &   $15$& $g=(\textbf{1}^{30}\textbf{3}^{29}\textbf{2}^{28}\textbf{1}^{27}\textbf{3}^{26}\textbf{2}^{25}
  \textbf{1}^{24}\textbf{2}^{22}\textbf{3}^{21}\textbf{3}^{19}\textbf{3}^{16}\textbf{3}^{15}
  \textbf{2}^{14}\textbf{2}^{13}\textbf{2}^{12}\textbf{3}^{11}\textbf{3}^{10}\textbf{2}^{9}
  \textbf{3}^{8}\textbf{2}^{7}\textbf{2}^{6}\textbf{1}^{5}\textbf{1}^{4}\textbf{1}^{3}
  \textbf{3}^{2}\textbf{1}^{0})$   \\
   \cline{2-4}
   &  $[[85,33]]$ &   $13$ & $g=(\textbf{1}^{26} \textbf{1}^{25}\textbf{3}^{24}\textbf{3}^{23}\textbf{2}^{22}\textbf{3}^{21}\textbf{3}^{20}\textbf{1}^{19}\textbf{2}^{18}\textbf{2}^{17}\textbf{2}^{16}\textbf{3}^{14}\textbf{2}^{13}\textbf{1}^{11} \textbf{1}^{9}\textbf{2}^{7} \textbf{2}^{6}\textbf{2}^{4}\textbf{2}^{3}\textbf{1}^{0})$   \\
    \cline{2-4}
    &  $[[85,37]]$ &   $12$ & $g=(\textbf{1}^{24} \textbf{1}^{23}\textbf{2}^{22}\textbf{3}^{21}\textbf{1}^{19}\textbf{1}^{18}\textbf{3}^{17}\textbf{3}^{16}\textbf{1}^{15}\textbf{1}^{14}\textbf{2}^{12}\textbf{1}^{10}\textbf{2}^{8}\textbf{1}^{7} \textbf{1}^{5}\textbf{2}^{4} \textbf{3}^{3}\textbf{2}^{1}\textbf{1}^{0})$   \\
     \cline{2-4}
     &  $[[85,41]]$ &   $11$ & $g=(\textbf{1}^{22} \textbf{3}^{21}\textbf{3}^{20}\textbf{2}^{19}\textbf{3}^{18}
     \textbf{3}^{16}\textbf{1}^{14}\textbf{3}^{13}\textbf{1}^{11}\textbf{3}^{9}
     \textbf{3}^{6}\textbf{1}^{5}\textbf{2}^{3}\textbf{1}^{2}  \textbf{1}^{1}\textbf{1}^{0})$   \\
      \cline{2-4}
     &  $[[85,45]]$ &   $10$ & $g=(\textbf{1}^{20} \textbf{3}^{19}\textbf{2}^{12}\textbf{1}^{10}\textbf{2}^{9}\textbf{3}^{8}
     \textbf{2}^{7}\textbf{2}^{5}\textbf{3}^{4}\textbf{2}^{3}\textbf{3}^{2} \textbf{1}^{1}\textbf{1}^{0})$   \\
      \cline{2-4}
  &  $[[85,49]]$ &   $9$ & $g=(\textbf{1}^{18} \textbf{3}^{17}\textbf{2}^{16}\textbf{1}^{15}\textbf{1}^{14}\textbf{3}^{13}\textbf{3}^{12}
  \textbf{1}^{11}\textbf{2}^{10}\textbf{1}^{9}\textbf{2}^{8}\textbf{3}^{7}\textbf{3}^{6}
  \textbf{3}^{5}\textbf{2}^{4}\textbf{1}^{3}\textbf{3}^{1}\textbf{1}^{0})$   \\
   \cline{2-4}
&  $[[85,53]]$ &   $8$ & $g=(\textbf{1}^{16} \textbf{3}^{15}\textbf{2}^{14}\textbf{1}^{11}\textbf{1}^{10}\textbf{3}^{9}
\textbf{1}^{8}\textbf{1}^{7}\textbf{2}^{6}\textbf{3}^{5}\textbf{3}^{4} \textbf{3}^{2}  \textbf{1}^{0})$   \\
 \cline{2-4}
 &  $[[91,7]]$ &   $21$ & $g=(\textbf{1}^{42} \textbf{3}^{35}\textbf{2}^{21}\textbf{3}^{7}   \textbf{1}^{0})$   \\
  \cline{2-4}
 &  $[[93,13]]$ &   $20$ &  {$g_1=(\textbf{1}^{6}\textbf{1}^4\textbf{1}^1\textbf{1}^0)$,
 $g_2=(\textbf{1}^{6}\textbf{1}^4\textbf{1}^2\textbf{1}^1\textbf{1}^0)$}   \\
 \cline{2-4}
 &  $[[93,33]]$ &   $15$ &  $g=(\textbf{1}^{30}  \textbf{2}^{29} \textbf{1}^{28}  \textbf{2}^{27}  \textbf{1}^{26} \textbf{2}^{24}   \textbf{3}^{21} \textbf{1}^{20}   \textbf{1}^{19} \textbf{1}^{18} \textbf{2}^{17}  \textbf{1}^{15} \textbf{1}^{13}     \textbf{1}^{12} \textbf{2}^{11} \textbf{3}^{10}  \textbf{3}^{9}   \textbf{2}^{6}   \textbf{3}^{4}  \textbf{2}^{3}  \textbf{2}^{2}  \textbf{1}^{1}   \textbf{1}^{0})$  \\
   \cline{2-4}
 &  $[[95,19]]$ &   $19$ & $g=(\textbf{1}^{38} \textbf{3}^{19}   \textbf{1}^{0})$   \\

\hline
\multirow{1}{*}{$1$} &  $[[45,8]]$  &   $9$&  {$g_1=(\textbf{1}^{19}\textbf{1}^{18}\textbf{1}^{16}\textbf{1}^{12}\textbf{1}^{10}\textbf{1}^{9}\textbf{1}^{6}\textbf{1}^{4}\textbf{1}^3\textbf{1}^0)$,
 $g_2=(\textbf{1}^{18}\textbf{1}^{15}\textbf{1}^{12}\textbf{1}^9\textbf{1}^0)$} \\

\hline
\multirow{20}{*}{$2$} &  $[[21,9]]$  &   $3$& {$g_1=(\textbf{1}^{6}\textbf{1}^4\textbf{1}^1\textbf{1}^0)$,
 $g_2=(\textbf{1}^{6}\textbf{1}^4\textbf{1}^2\textbf{1}^1\textbf{1}^0)$} \\
  \cline{2-4}
  &  $[[23,1]]$ &   $5$& $g=(\textbf{1}^{11}\textbf{1}^9\textbf{1}^7\textbf{1}^6\textbf{1}^5\textbf{1}^1\textbf{1}^0)$
  \\
   \cline{2-4}
  &  $[[31,1]]$ &   $7$& $g=(\textbf{1}^{15}\textbf{1}^{14}\textbf{1}^{13}\textbf{1}^{9}
  \textbf{1}^8\textbf{1}^3\textbf{1}^0)$
  \\
  \cline{2-4}
  &  $[[35,25]]$ &   $2$& $g=(\textbf{1}^{5}\textbf{2}^{4}\textbf{3}^{2}\textbf{2}^{1}\textbf{1}^0)$
  \\
   \cline{2-4}
  &  $[[35,17]]$ &   $4$& $g=(\textbf{1}^{9}\textbf{3}^{7}\textbf{3}^{6}\textbf{3}^{5}\textbf{3}^{4}\textbf{2}^{3}\textbf{2}^{2}\textbf{2}^1\textbf{1}^0)$
  \\
     \cline{2-4}
  &  $[[35,13]]$ &   $5$& $g=(\textbf{1}^{11}\textbf{3}^{10}\textbf{2}^{9}\textbf{1}^{8}\textbf{2}^{7}\textbf{2}^{6}\textbf{3}^{5}
\textbf{1}^{3}\textbf{2}^{2}\textbf{1}^1\textbf{1}^0)$
  \\
    \cline{2-4}
  &  $[[35,5]]$ &   $7$& $g=(\textbf{1}^{15}\textbf{1}^{14}\textbf{1}^{13}\textbf{1}^{12}\textbf{1}^{10}
  \textbf{1}^8\textbf{1}^6\textbf{1}^5\textbf{1}^4\textbf{1}^1\textbf{1}^0)$
  \\
     \cline{2-4}
  &  $[[47,1]]$ &   $11$& $g=(\textbf{1}^{23}\textbf{1}^{19}\textbf{1}^{18}\textbf{1}^{14}\textbf{1}^{13}\textbf{1}^{12}\textbf{1}^{10}
 \textbf{1}^{9} \textbf{1}^7\textbf{1}^6\textbf{1}^5\textbf{1}^{3}\textbf{1}^2\textbf{1}^1\textbf{1}^0)$
  \\
       \cline{2-4}
  &  $[[63,9]]$ &   $13$& $g=(\textbf{1}^{27}\textbf{1}^{26}\textbf{2}^{24}\textbf{3}^{23}\textbf{2}^{22}\textbf{1}^{19}\textbf{1}^{18}\textbf{2}^{17}\textbf{2}^{14}\textbf{2}^{13}\textbf{2}^{12}\textbf{2}^{11}\textbf{3}^{10}
 \textbf{3}^{9} \textbf{1}^{8}\textbf{3}^7\textbf{1}^6\textbf{1}^5\textbf{3}^{4}\textbf{2}^{3}\textbf{1}^2\textbf{1}^1\textbf{1}^0)$
  \\
       \cline{2-4}
  &  $[[63,21]]$ &   $10$& $g=(\textbf{1}^{21}\textbf{2}^{20}\textbf{3}^{19}\textbf{1}^{18}\textbf{3}^{17}\textbf{1}^{16}\textbf{3}^{15}\textbf{3}^{14}\textbf{3}^{13}\textbf{1}^{12}\textbf{3}^{11}\textbf{2}^{10}\textbf{3}^{8}
 \textbf{3}^{6} \textbf{2}^{4}\textbf{2}^2\textbf{1}^0)$
  \\
 \cline{2-4}
  &  $[[63,33]]$ &   $7$& $g=(\textbf{1}^{15}\textbf{1}^{13}\textbf{1}^{10}\textbf{2}^{9}\textbf{2}^{8}\textbf{2}^{7}\textbf{3}^{6}
 \textbf{2}^{5} \textbf{2}^{3}\textbf{3}^2\textbf{3}^1\textbf{3}^0)$
  \\
     \cline{2-4}
  &  $[[63,45]]$ &   $4$& $g=( \textbf{1}^{9}\textbf{3}^{7}\textbf{2}^{5}\textbf{2}^{4}
 \textbf{2}^{2} \textbf{3}^{1}\textbf{1}^0)$
  \\
  \cline{2-4}
  &  $[[71,1]]$ &   $17$& $g=(\textbf{1}^{35}\textbf{1}^{33}\textbf{1}^{28}\textbf{1}^{27}\textbf{1}^{26}\textbf{1}^{25}\textbf{1}^{24}\textbf{1}^{17}\textbf{1}^{13}\textbf{1}^{8} \textbf{1}^{7}\textbf{1}^{5}\textbf{1}^{4}\textbf{1}^{1}\textbf{1}^0)$
  \\
  \cline{2-4}
  &  $[[77,47]]$ &   $7$& $g=(\textbf{1}^{15}\textbf{2}^{13}\textbf{3}^{12}\textbf{1}^{11}\textbf{2}^{10}\textbf{1}^{7}\textbf{1}^{6}\textbf{2}^{5}\textbf{2}^{4}\textbf{3}^{3} \textbf{1}^{2}\textbf{1}^{1} \textbf{1}^0)$
  \\
   \cline{2-4}
  &  $[[91,13]]$ &   $19$& $g=(\textbf{1}^{39}\textbf{1}^{38}\textbf{2}^{36}\textbf{2}^{35}\textbf{3}^{34}\textbf{3}^{32}\textbf{1}^{31}\textbf{3}^{30}\textbf{2}^{29}\textbf{2}^{27}\textbf{3}^{26}\textbf{1}^{25}\textbf{2}^{24}\textbf{3}^{23}\textbf{3}^{22}\textbf{2}^{19}\textbf{3}^{18}\textbf{3}^{17}\textbf{2}^{16}\textbf{3}^{15}\textbf{2}^{14}\textbf{1}^{13}\textbf{1}^{9}\textbf{3}^{8}\textbf{2}^{7}\textbf{3}^{6}\textbf{1}^{4}\textbf{2}^{3}\textbf{2}^{2} \textbf{1}^0)$
  \\
  \cline{2-4}
  &  $[[91,25]]$ &   $16$& $g=(\textbf{1}^{33}\textbf{2}^{32}\textbf{3}^{30}\textbf{2}^{28}\textbf{1}^{27}\textbf{1}^{26}\textbf{2}^{25}\textbf{1}^{24}\textbf{1}^{23}\textbf{1}^{21}\textbf{1}^{19}\textbf{2}^{18}\textbf{1}^{17}\textbf{3}^{16}\textbf{1}^{14}\textbf{2}^{13}\textbf{1}^{12}\textbf{2}^{11}\textbf{2}^{16}\textbf{3}^{15}\textbf{2}^{14}\textbf{1}^{10}\textbf{2}^{9}\textbf{3}^{8}\textbf{1}^{7}\textbf{1}^{5}\textbf{1}^{3}\textbf{1}^{2}\textbf{2}^{1} \textbf{1}^0)$
  \\
 \cline{2-4}
  &  $[[91,37]]$ &   $13$& $g=(\textbf{1}^{27}\textbf{3}^{26}\textbf{3}^{25}\textbf{3}^{24}\textbf{2}^{23}\textbf{2}^{22}\textbf{3}^{21}\textbf{3}^{20}\textbf{3}^{19}\textbf{1}^{18}\textbf{1}^{17}\textbf{2}^{16}\textbf{2}^{14}\textbf{2}^{11}\textbf{2}^{10}\textbf{1}^{9}\textbf{3}^{8}\textbf{1}^{7}\textbf{1}^{6}\textbf{3}^{5}\textbf{3}^{4}\textbf{1}^{3}\textbf{2}^{2}  \textbf{1}^0)$
    \\
 \cline{2-4}
  &  $[[91,49]]$ &   $10$& $g=(\textbf{1}^{21}\textbf{2}^{19}\textbf{1}^{15}\textbf{3}^{14}\textbf{3}^{13}\textbf{1}^{12}\textbf{1}^{11}\textbf{2}^{9}\textbf{2}^{5}\textbf{1}^{4}\textbf{1}^{3}\textbf{2}^{1} \textbf{1}^0)$
  \\
 \cline{2-4}
  &  $[[91,61]]$ &   $7$& $g=(\textbf{1}^{15}\textbf{1}^{13}\textbf{2}^{12}\textbf{2}^{11}\textbf{1}^{10}\textbf{1}^{9}\textbf{1}^{8}\textbf{3}^{7}\textbf{3}^{5}\textbf{3}^{2}\textbf{2}^{1}\textbf{2}^{1} \textbf{1}^0)$
  \\
 \cline{2-4}
  &  $[[91,73]]$ &   $4$& $g=(\textbf{1}^{9}\textbf{1}^{8}\textbf{3}^{5}\textbf{3}^{4}\textbf{1}^{3}\textbf{3}^{2} \textbf{1}^{1} \textbf{1}^0)$
  \\

\hline
\end{tabular}
\end{table*}


\begin{theorems}
\label{theoremquantumburst1}
Let ${C}=[n,k]_{q^2}$ be a classical cyclic code  with a parity check matrix   $\mathbf{H}$ such that $\mathbf{H}\mathbf{H}^\dag=0$. Denote the generator matrix of ${C}$ by  $\mathbf{G}$. Let $1\leq \ell\leq r$ and let $A_{\imath}=[\imath,\imath+\ell-1]$, where $1\leq \imath\leq n-2\ell+1$.  Let $\mathcal{L}$ be the
largest integer $\ell$  such that exactly one of the following two terms is satisfied
\begin{itemize}
\item[1)]  Every $\mathcal{M}^{(\ell)}_{A_{\imath}}$ is of full rank for all $1\leq \imath\leq n-2\ell+1$.

\item[2)] For each $\imath\in [1, n-2\ell+1]$,   if $\mathcal{M}^{(\ell)}_{A_{\imath}}$   is not of full rank, then     the condition
$\mathbf{G}^\dagger e^T=\mathbf{G}^\dagger f^T$ holds for all $(e,f)\in\boxplus_{\mathcal{M}^{(\ell)}_{A_{\imath}}}$.
\end{itemize}
There exists an $[[n,2k-n]]$ QCC  ${Q}$ which can correct any quantum burst error of length $\mathcal{L}$ or less.
\end{theorems}

\begin{IEEEproof}
The proof is given in Appendix \ref{appendixtheoremquantumburst1}.
\end{IEEEproof}

In order to determine the burst error correction limit of QBECCs, we need to verify whether  each $\mathcal{M}^{(\ell)}_{A_{\imath}}$ is of full rank for  $1\leq \imath\leq n-2\ell+1$ according to Theorem \ref{theoremquantumburst1}. If some $\mathcal{M}^{(\ell)}_{A_{\imath}}$ is not of full rank, we need to find the set $ \boxplus_{\mathcal{M}^{(\ell)}_{A_{\imath}}}$ to verify whether the corresponding burst errors are degenerate or not.  Recall that we can transform $\mathcal{M}^{(\ell)}_{A_{\imath}}$ to a diagonal matrix with form in Eq.   (\ref{diagonalmatrixform}) according to Lemma \ref{diagonalmatrixformlemma}. We can conduct     Gaussian elimination by rows to realize the transformation. Then the set $\boxplus_{\mathcal{M}^{(\ell)}_{A_{\imath}}}$ can be derived directly by the ``$X$" part in  Eq.   (\ref{diagonalmatrixform}).

In Algorithm \ref{BurstLengthOfQCCs}, we present the algorithm for determining the burst error correction limit of QCCs. Now we analyze the time complexity of   Algorithm \ref{BurstLengthOfQCCs}. In order to facilitate the understanding, we only   give an upper bound to the time complexity of Algorithm \ref{BurstLengthOfQCCs}, rather than  the exact time complexity of it.   From line 1 to line 4, the algorithm verifies whether the cyclic code satisfies the dual containing restrict and the time complexity is $O(r^2n)$. From line 6 to line 20, the algorithm determines whether the QBECC can correct all burst errors of length $1$.  The time complexity is $O(rn^2)$. Line 21 to line 37 are the main part of Algorithm \ref{BurstLengthOfQCCs}. In each subcycle, the time complexity of the rows Gaussian elimination  is related to the burst error correction ability $\ell$. The upper bound of the time complexity of the  rows Gaussian elimination is $O(r^3)$. Thus the total time complexity from line 21 to line 37 is $O(r^4n)$. Overall, the time complexity of Algorithm  \ref{BurstLengthOfQCCs} is $O(r^4n+rn^2)$.

In Table \ref{Computer Searching for QBECCs with Alg1} and Table \ref{Computer Searching for degenerate QBECCs with Alg1}, we list the   optimal and nearly optimal QCCs of length $n<100$ by using Algorithm  \ref{BurstLengthOfQCCs} to compute their burst error correction abilities. Denote the parameters of   QCCs by $Q=[[n,k;\mathcal{L}]]$, and   denote by $\Delta=n-k-4\mathcal{L}$. In Table \ref{Computer Searching for QBECCs with Alg1},  we list optimal or nearly optimal nondegenerate QCCs. While in Table \ref{Computer Searching for degenerate QBECCs with Alg1}, we list optimal or nearly optimal degenerate QCCs which have better burst error correction abilities than any nondegenerate QCCs constructed from the CSS or the Hermitian constructions.
In Table \ref{Computer Searching for QBECCs with Alg1} and Table \ref{Computer Searching for degenerate QBECCs with Alg1}, the bold numbers ``$\textbf{1}-\textbf{3}$'' in subscripts and the  numbers in superscripts of the generator polynomials stand for the
coefficients and the exponents, respectively.
  In   Table \ref{Computer Searching for degenerate QBECCs with Alg1}, we denote $\ell_0$ by the nondegenerate burst error correction ability of the $[[n,k;\mathcal{L}]]$ QCC.  In this paper, we run all the algorithms in Magma software  (V2.28-3). The operating system is Ubuntu 22.04 LTS and the processor is Intel i5-12490F.

\begin{algorithm}
\caption{The Burst Error Correction Limit of QCCs.}
\label{BurstLengthOfQCCs}
\begin{algorithmic}[1]
\Require
   $\textbf{H}$, $\textbf{G}$;
\Ensure The burst error correction limit  $\mathcal{L}$.
\If{$\mathbf{H}\mathbf{H}^\dag\ne0$}
\State$//$ \verb"Fail to construct a QCC".
\State return null;
\EndIf
\State Initialization: $ r = \textrm{rank}(\mathbf{H}),\ell=1$;
\For{$i\in[1,n]$}
\If{$\mathbf{H}(;i)=0$}
\State$//$ The limit $\mathcal{L}=0$.
\State return 0;
\EndIf
\EndFor
\For{$i\in[1,n-1]$}
\For{$j\in[i+1,n]$}
\State $S_H = \mathbf{H}(;i)+\mathbf{H}(;j)$, $S_{G^\dagger} = \mathbf{G}^\dagger(;i)+\mathbf{G}^\dagger(;j)$;
\If{$S_H=0 \mod q$ and $S_{G^\dagger} \ne0 \mod q$}
\State$//$ The limit $\mathcal{L}=0$.
\State return 0;
\EndIf
\EndFor
\EndFor
\While{$\ell\leq r /2$}
\For{ $\imath\in [1,n-2\ell+1]$}
\State $A_{\imath}=[\imath,\imath+\ell-1];$
\State$//$ \verb"Do Gaussian elimination to" $\mathcal{M}^{(\ell)}_{A_{\imath}}$.
\State  $\widetilde{\mathcal{M}}^{(\ell)}_{A_{\imath}}=$ RowsGaussianElimination$(\mathcal{M}^{(\ell)}_{A_{\imath}})$;
\If{rank$(\widetilde{\mathcal{M}}^{(\ell)}_{A_{\imath}})<\ell$}
\State  $\boxplus_{\mathcal{M}^{(\ell)}_{A_{\imath}}}=\{(e_1,f_1),\cdots,(e_v,f_v)\}$;
\For{$ j\in [1,v]$}
\If{$\mathbf{G}^\dagger e_j^T\ne\mathbf{G}^\dagger f_j^T$}
\State$//$ $e$ \verb"and" $f$ are \verb"nondegenerate".
\State  return $\ell$;
\EndIf
\EndFor
\EndIf
\EndFor
\State $\ell=\ell+1$;
\EndWhile
\State return $\mathcal{L}=\ell$;
\end{algorithmic}
\end{algorithm}

\begin{table*}
\caption{Computer searching for optimal and nearly optimal degenerate  QCCs of length $n<100$.   }
\label{Computer Searching for degenerate QBECCs with Alg1}
\centering
\begin{tabular}{|l|l|l|l|l|l| }
\hline
 $\Delta$&  $[[n,k]]$&$\mathcal{L}$&$\ell_0$&Generator Polynomials
 \\
 \hline
 \multirow{17}{*}{$0$} &   $[[25,1]]$&   $6$&$5$ & $g=(\textbf{1}^{12}\textbf{2}^{11}\textbf{1}^{10}\textbf{2}^7\textbf{3}^6\textbf{2}^5\textbf{1}^2\textbf{2}^1\textbf{1}^0)$   \\
   \cline{2-5}
  &  $[[29,1]]$ &   $7$&$6$& $g=(\textbf{1}^{14}\textbf{2}^{13}\textbf{2}^{11}\textbf{3}^{10}\textbf{1}^{9}\textbf{3}^{8}
  \textbf{2}^7\textbf{3}^6\textbf{1}^5\textbf{3}^{4}\textbf{2}^{ 3 }\textbf{2}^1\textbf{1}^0)$   \\
   \cline{2-5}
  &  $[[37,1]]$ &   $9$& $8$ & $g=(\textbf{1}^{18}\textbf{2}^{17}\textbf{1}^{16}\textbf{1}^{15}\textbf{2}^{14}\textbf{2}^{13}
\textbf{3}^{12}\textbf{1}^{11}\textbf{2}^{10}\textbf{1}^{9}\textbf{2}^{8}\textbf{1}^7\textbf{3}^{6}\textbf{2}^{5}\textbf{2}^4\textbf{1}^{3}\textbf{1}^{2}\textbf{2}^1\textbf{1}^0)$ \\
   \cline{2-5}
  &  $[[41,1]]$ &   $10$&$9$ & $g=(\textbf{1}^{20}\textbf{2}^{19}\textbf{1}^{18}\textbf{1}^{17}\textbf{3}^{16}\textbf{3}^{14}\textbf{2}^{13}\textbf{2}^{12}
 \textbf{2}^{11}\textbf{3}^{10}\textbf{2}^{9}\textbf{2}^{8}\textbf{2}^7\textbf{3}^{6}\textbf{3}^4\textbf{1}^{3}\textbf{1}^{2}\textbf{2}^1\textbf{1}^0)$ \\
   \cline{2-5}
  &  $[[53,1]]$ &   $13$&$12$&  $g=(\textbf{1}^{26}\textbf{2}^{25}\textbf{1}^{24}\textbf{1}^{22}\textbf{1}^{21}\textbf{2}^{19} \textbf{1}^{17}\textbf{3}^{16}\textbf{2}^{15}\textbf{3}^{14}\textbf{3}^{13}\textbf{3}^{12}
 \textbf{2}^{11}\textbf{3}^{10}\textbf{1}^{9} \textbf{2}^7\textbf{1}^{5}\textbf{1}^4 \textbf{1}^{2}\textbf{2}^1\textbf{1}^0)$ \\
   \cline{2-5}
  &  $[[61,1]]$ &   $15$& $14$ & $g=(\textbf{1}^{30}\textbf{2}^{29}\textbf{3}^{27}\textbf{1}^{26}\textbf{1}^{25}\textbf{3}^{24}\textbf{3}^{23}
\textbf{1}^{22}\textbf{2}^{20}\textbf{1}^{19} \textbf{1}^{18}\textbf{2}^{17}\textbf{3}^{16}\textbf{3}^{15}\textbf{3}^{14}\textbf{2}^{13}\textbf{1}^{12}
 \textbf{1}^{11}\textbf{2}^{10}\textbf{1}^{8} \textbf{3}^7\textbf{3}^{6}\textbf{1}^{5}\textbf{1}^4\textbf{3}^{3} \textbf{2}^1\textbf{1}^0)$ \\
   \cline{2-5}
  &  $[[65,1]]$ &   $16$& $15$ &  $g=(\textbf{1}^{32}\textbf{2}^{31}\textbf{1}^{30}\textbf{1}^{29}\textbf{3}^{28}\textbf{1}^{27}
\textbf{3}^{21}\textbf{1}^{20}\textbf{2}^{19} \textbf{1}^{17}\textbf{2}^{16}\textbf{1}^{15} \textbf{2}^{13}\textbf{1}^{12}
 \textbf{3}^{11} \textbf{1}^{5}\textbf{3}^4\textbf{1}^{3} \textbf{1}^{2}\textbf{2}^1\textbf{1}^0)$ \\
   \cline{2-5}
 & $[[73,1]]$&$18$&$16$ &\tabincell{l}{$g_1=(\textbf{1}^{36}\textbf{1}^{35}\textbf{1}^{34}\textbf{1}^{31}\textbf{1}^{29}\textbf{1}^{28}\textbf{1}^{27}\textbf{1}^{21}\textbf{1}^{19}\textbf{1}^{18}\textbf{1}^{17}\textbf{1}^{15}\textbf{1}^{9}\textbf{1}^{8}\textbf{1}^{7}\textbf{1}^{5}\textbf{1}^2\textbf{1}^1\textbf{1}^0)$,
\\$g_2=(\textbf{1}^{36}\textbf{1}^{33}\textbf{1}^{31}\textbf{1}^{29}\textbf{1}^{27}\textbf{1}^{25}\textbf{1}^{24}\textbf{1}^{22}\textbf{1}^{20}\textbf{1}^{19}\textbf{1}^{18}\textbf{1}^{17}\textbf{1}^{16}\textbf{1}^{14}\textbf{1}^{12}\textbf{1}^{11}\textbf{1}^{9}\textbf{1}^{7}\textbf{1}^{5}\textbf{1}^{3}\textbf{1}^0)$}  \\
      \cline{2-5}
  &  $[[75,3]]$&$18$&$15$ &$g=(\textbf{1}^{36}\textbf{2}^{33}\textbf{1}^{30}\textbf{2}^{21}\textbf{3}^{18}\textbf{2}^{15}
\textbf{1}^{6}\textbf{2}^{3} \textbf{1}^0)$ \\
     \cline{2-5}
  &  $[[85,1]]$&$21$&$20$ &$g=(\textbf{1}^{42}\textbf{3}^{41}\textbf{2}^{39}\textbf{1}^{38}\textbf{1}^{37}\textbf{2}^{36}\textbf{3}^{35}
\textbf{3}^{34}\textbf{3}^{33}\textbf{1}^{31}\textbf{1}^{30}\textbf{3}^{29}
\textbf{3}^{28}\textbf{2}^{27}\textbf{1}^{23}\textbf{1}^{22}\textbf{3}^{21}\textbf{1}^{20}
\textbf{1}^{19}\textbf{2}^{15} \textbf{3}^{14}\textbf{3}^{13}\textbf{1}^{12}\textbf{1}^{11}\textbf{3}^{9}\textbf{3}^{8}\textbf{3}^{7}\textbf{2}^{6}
\textbf{1}^{5}
\textbf{1}^{4}\textbf{2}^{3}\textbf{3}^{1} \textbf{1}^0)$ \\
     \cline{2-5}
  &  $[[87,3]]$&$21$&$18$ &$g=(\textbf{1}^{42} \textbf{3}^{39}\textbf{3}^{33}
\textbf{2}^{30}\textbf{1}^{27}
\textbf{2}^{24} \textbf{3}^{21}
\textbf{2}^{18}\textbf{1}^{15} \textbf{2}^{12} \textbf{3}^{9} \textbf{3}^{3}  \textbf{1}^0)$ \\
  \cline{2-5}
  &  $[[89,1]]$&$22$&$20$ &\tabincell{l}{$g_1=(\textbf{1}^{44}\textbf{1}^{39}\textbf{1}^{35}\textbf{1}^{34}\textbf{1}^{32}\textbf{1}^{31}\textbf{1}^{30}\textbf{1}^{29}\textbf{1}^{22}\textbf{1}^{15}\textbf{1}^{14}\textbf{1}^{13}\textbf{1}^{12}\textbf{1}^{10}\textbf{1}^{9}\textbf{1}^{5}\textbf{1}^0)$,
\\$g_2=(\textbf{1}^{44}\textbf{1}^{43}\textbf{1}^{42}\textbf{1}^{41}\textbf{1}^{40}\textbf{1}^{35}\textbf{1}^{34}\textbf{1}^{33}\textbf{1}^{20}\textbf{1}^{19}\textbf{1}^{18}\textbf{1}^{31}\textbf{1}^{26}\textbf{1}^{24}\textbf{1}^{23}\textbf{1}^{22}\textbf{1}^{21}\textbf{1}^{20}\textbf{1}^{18}\textbf{1}^{13}\textbf{1}^{11}\textbf{1}^{10}\textbf{1}^{9}\textbf{1}^{4}\textbf{1}^{3}\textbf{1}^{2}\textbf{1}^{1}\textbf{1}^0)$}  \\
    \cline{2-5}
  &  $[[97,1]]$&$24$&$23$ &$g=(\textbf{1}^{48}\textbf{3}^{47}\textbf{1}^{46}\textbf{2}^{43}\textbf{1}^{42}\textbf{3}^{41}\textbf{2}^{40}
\textbf{2}^{39} \textbf{2}^{37} \textbf{2}^{35}
\textbf{1}^{34}\textbf{2}^{33}\textbf{3}^{31}\textbf{2}^{30}\textbf{2}^{29}
\textbf{3}^{26}\textbf{3}^{25}\textbf{2}^{24}\textbf{3}^{23}\textbf{3}^{22}\textbf{2}^{19}\textbf{2}^{18}
\textbf{3}^{17}\textbf{2}^{15} \textbf{1}^{14}\textbf{2}^{13}\textbf{2}^{11} \textbf{2}^{9}\textbf{2}^{8}\textbf{3}^{7}\textbf{1}^{6}
\textbf{2}^{5}
\textbf{1}^{2} \textbf{3}^{1} \textbf{1}^0)$ \\

\hline
\multirow{2}{*}{$1$} &  $[[51,2]]$  &   $12$&$9$&{$g_1=(\textbf{1}^{25}\textbf{1}^{24}\textbf{1}^{16}\textbf{1}^{15}\textbf{1}^{13}\textbf{1}^{12}\textbf{1}^{10}\textbf{1}^{9}\textbf{1}^{1}\textbf{1}^0)$,
 $g_2=(\textbf{1}^{24}\textbf{1}^{21}\textbf{1}^{18}\textbf{1}^{12}\textbf{1}^{6}\textbf{1}^{3}\textbf{1}^0)$} \\
  \cline{2-5}
  &  $[[85,4]]$&$20$&$15$ & {$g_1=(\textbf{1}^{41} \textbf{1}^{40}\textbf{1}^{26}\textbf{1}^{25}\textbf{1}^{21}\textbf{1}^{20}\textbf{1}^{16}  \textbf{1}^{15}\textbf{1}^{1}\textbf{1}^0)$,
 $g_2=(\textbf{1}^{40}\textbf{1}^{35}\textbf{1}^{30}\textbf{1}^{20}\textbf{1}^{10}\textbf{1}^{5}\textbf{1}^0)$}
  \\

\hline
\multirow{3}{*}{$2$} &  $[[35,1]]$  &   $8$&$7$& $g=(\textbf{1}^{17}\textbf{1}^{15}\textbf{1}^{14}\textbf{3}^{10}\textbf{3}^{8}\textbf{3}^7\textbf{1}^3\textbf{1}^1\textbf{1}^0)$ \\
  \cline{2-5}
 & $[[79,1]]$&$19$&$18$ &$g=(\textbf{1}^{39}\textbf{1}^{36}\textbf{1}^{35}\textbf{1}^{31}\textbf{1}^{30}\textbf{1}^{29}
\textbf{1}^{27}\textbf{1}^{26}\textbf{1}^{25}\textbf{1}^{24}\textbf{1}^{21}\textbf{1}^{20}
\textbf{1}^{19}\textbf{1}^{18}\textbf{1}^{16}\textbf{1}^{14}\textbf{1}^{13}\textbf{1}^{11}\textbf{1}^{5}
\textbf{1}^{4}\textbf{1}^{2}\textbf{1}^1\textbf{1}^0)$ \\
   \cline{2-5}
  &  $[[93,3]]$&$22$&$20$ &$g=(\textbf{1}^{45} \textbf{1}^{44}\textbf{1}^{43}\textbf{1}^{41}\textbf{1}^{40}\textbf{1}^{39}\textbf{1}^{38}  \textbf{1}^{37}\textbf{1}^{34}\textbf{1}^{33}\textbf{1}^{32}\textbf{1}^{29}\textbf{1}^{28}\textbf{1}^{25}\textbf{1}^{23}\textbf{1}^{20}\textbf{1}^{19}\textbf{1}^{17}\textbf{1}^{15}\textbf{1}^{13}\textbf{1}^{11}\textbf{1}^{10}\textbf{1}^{6}\textbf{1}^{5}\textbf{1}^{4}\textbf{1}^{1}\textbf{1}^0)$ \\
\hline
\end{tabular}
\end{table*}

\section{The True Burst Error Correction Ability of Quantum Reed-Solomon Codes}
\label{polynomialalgorithmburstquantumRScodes}
As with classical RS codes \cite{lin2004error}, quantum RS codes are effective at correcting both quantum random errors \cite{ouyang2014concatenated, thommesen2003existence} and quantum burst errors \cite{grassl1999quantum}. Moreover, there exists a lower bound for the burst error correction ability of  quantum  RS codes \cite{qian2015constructions,lin2004error}. However, this lower bound is not tight and does not give the true burst error correction ability of quantum RS codes. In this section, we give a  polynomial-time algorithm to determine that.

Let ${C}_{RS}=[n=q^m-1,k]_{q^m}$ be a classical narrow sense RS code such that $n\leq 2k$. Then we know that ${C}_{RS}^\bot\subseteq {C}_{RS}$    and we can construct a ${Q}_{RS}=[[n,2k-n]]_{q^m}$ quantum RS code by using the CSS construction \cite{rotteler2004quantum}. Let $[{C}_{RS}]=[mn,mk]_q$ be the $q$-ary image of the  RS code $ {C}_{RS}$. We use the self-dual basis of $\mathbb{F}_{q^m}$ over $\mathbb{F}_q$ so that the dual-containing relationship can be maintained in the $q$-ary extension of  ${C}_{RS}$ \cite{macwilliams1977theory,cheng2021towards}.
\begin{lemmas}{\cite{ashikhmin2001asymptotically}}
\label{dualbinaryexpansion}
Let $\{\alpha_1,\cdots,\alpha_m\}$ be a self-dual basis of $\mathbb{F}_{q^m}$ over $\mathbb{F}_q$, i.e., $\textrm{Tr}(\alpha_i\alpha_j)=\delta_{ij}$ for all $1\leq i,j\leq m$.     Let $[ {C}_{RS}] $ and $[ {C}_{RS}]^\bot$ be the $q$-ary images of  $ {C}_{RS}$ and ${C}_{RS}^\bot$ under the basis $\{\alpha_1,\cdots,\alpha_m\}$, respectively. If ${C}_{RS}^\bot\subseteq {C}_{RS}$, then $[ {C}_{RS}]^\bot\subseteq [ {C}_{RS}]$ and $[ {C}_{RS}]^\bot$ is the dual  of $[ {C}_{RS}]$.
\end{lemmas}

Moreover, there exists the following relationship between the codeword of a RS code and that of its $q$-ary expansion.
\begin{lemmas}
\label{relationshipofRSandExpan}
Let  $C=[n,k]_{q^m}$ be  a $q^m$-ary RS code. Denote $\mathbf{D}=[C]$ by the $q$-ary image of $C$ under  a self-dual basis  $\{\alpha_1,\cdots,\alpha_m\}$. Let $\mu_1$ and $\mu_2$ be any  two codewords of   $C$.  Let $\mathbf{v}_1$ and $\mathbf{v}_2$ be    any two  codewords of $\mathbf{D}$. Map $\mathbf{v}_1$ and $\mathbf{v}_2$ to two vectors $\nu_1,\nu_2 \in\mathbb{F}_{q^m}^n$ by using the basis $\{\alpha_1,\cdots,\alpha_m\}$. If $\mu_1 \ne \mu_2$ and $\mathbf{v}_1 \ne \mathbf{v}_2$, then $[\mu_1] \ne [\mu_2]$ and  $\nu_1\ne\nu_2 $.
\end{lemmas}
\begin{IEEEproof}
The proof is given in Appendix \ref{appendixrelationshipofRSandExpan}.
\end{IEEEproof}

By using Lemma \ref{dualbinaryexpansion} and the CSS construction, we can construct a $[ {Q}_{RS}]=[[mn,2mk-mn]]_q$ quantum code and we call it the image of the quantum RS code ${Q}_{RS}$.
 Denote by $\hbar=\lfloor (n-k)/2\rfloor$.  Then the $q$-ary image code $[ {Q}_{RS}]$  can correct any quantum burst error of length $(\hbar-1)m+1$ or less \cite{lin2004error,qian2015constructions}.
 \begin{lemmas}[\cite{lin2004error,qian2015constructions}]
 \label{binaryexpanburstbound}
 Let $C=[n,k]_{q^m} $ be  a classical RS code such that $C^\bot\subseteq C$. There exists a quantum code with parameters $Q=[[nm,m(2k-n)]]_q$ which can correct any quantum burst error of length $lm+1$ or less, where $l=\lfloor (n-k)/2\rfloor-1$.
 \end{lemmas}

 However, the burst error correction ability of $[{Q}_{RS}]$ is  a lower bound which  does not
give the true burst error correction ability of $[{Q}_{RS}]$. In this work, we give a polynomial-time algorithm to calculate the true burst error correction ability of quantum RS codes.

Let $B_{\imath}=[\imath,\imath+\hbar-1]$, where $1\leq \imath\leq n-2\hbar-1$.
  Let  $\mathcal{M}^{(\hbar+1)} $ be the $(r-\hbar-1)\times(n-\hbar-1)$  matrix formed by deleting
the last $\hbar+1$ rows and   last $\hbar+1$ columns of the parity-check
matrix $\mathbf{H}$ of the RS code.
Let $\mathcal{M}^{(\hbar+1)}_{B_\imath}$  be the subblock  of $\hbar+1$ consecutive columns  of $\mathcal{M}^{(\hbar+1)} $ indexed by $B_\imath$.
  Denote the set of columns of $\mathcal{M}^{(\hbar+1)}_{B_\imath}
$ by $\{\alpha_{B_\imath}^{(1)},\cdots,\alpha_{B_\imath}^{(\hbar+1)}\}$. We have the following result about $\mathcal{M}^{(\hbar+1)}_{B_\imath}
$.
\begin{lemmas}
\label{RSMBl}
The rank of each  $\mathcal{M}^{(\hbar+1)}_{B_\imath}$ satisfies  $\hbar-1\leq rank(\mathcal{M}^{(\hbar+1)}_{B_\imath}) \leq \hbar$, where $1\leq \imath\leq n-2\hbar-1$.
\end{lemmas}
\begin{IEEEproof}
The proof is given in Appendix \ref{appendixrelationshipofRSandExpan}.
\end{IEEEproof}

  Denote   a set of MLI columns in  $\mathcal{M}^{(\hbar+1)}_{B_\imath}$ by $\mathscr{D}_H=\{\alpha_1,\cdots,\alpha_{u}\}$, where $\alpha_u\in\mathcal{M}^{(\hbar+1)}_{B_\imath} $ and $  u $ is the rank of $\mathcal{M}^{(\hbar+1)}_{B_\imath}$.
According to Lemma \ref{RSMBl}, we have $\hbar-1\leq u\leq \hbar $. Let $\widehat{\mathscr{D}}_H= \{\beta_1,\cdots,\beta_v\}$  as the complementary set of $\mathscr{D}_H$, where $v=\hbar+1-u$.   Each element in $\widehat{\mathscr{D}}_H$ can be represented by a linear combination of vectors in $\mathscr{D}_H$, i.e., $\beta_i=a_{i1}\alpha_1+\cdots+a_{iu}\alpha_u$ for $1\leq i\leq v$. Denote by  a burst error  $e_i$ ($1\leq i\leq v$) whose nonzero components fall in   the positions of $\{\alpha_1,\cdots,\alpha_u,\beta_i\}$ in $\mathbf{H}$.
  Then there exists a burst error $f_i$ whose nonzero components falling entirely   in the last $\hbar$  positions such that $ e_i+f_i\in {C}$ for each $1\leq i\leq v$.  We define a set of $e_i$ and $f_i$ as
 \begin{equation}
   {\boxplus}_{\mathcal{M}^{(\hbar+1)}_{B_\imath}}=\{(e_1,f_1),\cdots,(e_v,f_v)\},
 \end{equation}
  where  $1\leq v\leq 2$, and

  \begin{eqnarray}
\nonumber
\boxtimes_{\mathcal{M}^{(\hbar+1)}_{B_\imath}}=\big\{\lambda_1A_1+\lambda_2A_2|\forall \lambda_1,\lambda_2\in \mathbb{F}_{q^m}, \hspace*{5mm} \\
 (\lambda_1,\lambda_2)\ne(0,0), \forall A_1,A_2 \in \boxplus_{\mathcal{M}^{(\hbar+1)}_{B_\imath}} \big\}.
\end{eqnarray}

 Furthermore, if $\mathbf{G}e^T=\mathbf{G}f^T$, then $e$ and $f$ are degenerate, otherwise, they are nondegenerate. Therefore we define
  \begin{equation}
   \widehat{\boxtimes}_{\mathcal{M}^{(\hbar+1)}_{B_\imath}}=\{(e,f)|(e,f)\in \boxtimes_{\mathcal{M}^{(\hbar+1)}_{B_\imath}},  \mathbf{G}e^T\ne\mathbf{G}f^T\}.
 \end{equation}
  Then we have the following result about the burst error correction limit of the image of quantum RS codes.

\begin{theorems}
\label{theoremQRSburstlimit}
Let ${C}=[n=q^m-1,k]_{q^m}$ be a classical RS code satisfying the dual containing relationship, i.e., ${C}^\bot\subseteq {C}$.   Let $\hbar=\lfloor (n-k)/2\rfloor$ and let  $B_{\imath}=[\imath,\imath+\hbar]$, where $1\leq \imath\leq n-2\hbar-1$. Let $\mathcal{L}$ be largest integer $\ell$ such that there does not exist $(e,f)\in  \widehat{\boxtimes}_{\mathcal{M}^{(\hbar+1)}_{B_\imath}}$ with
\begin{equation}\label{RSImageburstMax}
\max\{\textrm{bl}([e]),\textrm{bl}([f])\}\leq \ell.
\end{equation}
 
There exists a $[[n,2k-n]]_{q^m}$ quantum RS code $Q$ which  can correct any quantum burst error of length $\mathcal{L} $ or less.
\end{theorems}
\begin{IEEEproof}
The proof is given in Appendix \ref{appendixrelationshipofRSandExpan}.
\end{IEEEproof}

In Algorithm \ref{BurstLOfQRS}, we present the algorithm for determining the burst error correction limit of quantum RS codes. Similar to the analysis  of the time complexity of   Algorithm \ref{BurstLengthOfQCCs}, the time complexity of Algorithm \ref{BurstLOfQRS} is $O(n\hbar^3+mn^3)$.
In Table \ref{Computer Searching for QBECCRSs}, we compute the true burst error correction ability of several quantum RS codes  by using Algorithm  \ref{BurstLengthOfQCCs}.  Denote the parameters of quantum RS codes by $Q=[[n,k]]_{2^m}$. Denote by $\mathcal{L}$ the true burst error correction ability computed by using Algorithm \ref{BurstLOfQRS}, and  denote by $\ell_L$ the lower bound in \cite{qian2015constructions,lin2004error}. We show that the true burst error correction abilities of quantum RS codes  are better   than the lower bound in \cite{qian2015constructions,lin2004error}.


\begin{algorithm}
\caption{The True Burst Error Correction Limit of Quantum Reed-Solomon Codes.}
\label{BurstLOfQRS}
\begin{algorithmic}[1]
\Require
   $\textbf{H}$, $\textbf{G}$ of a RS code $C$;
\Ensure The burst error correction limit of the quantum RS code $Q$.
\If{$\mathbf{H}\mathbf{H}^\dag\ne0$}
\State$//$ \verb"Fail to construct a quantum RS code".
\State return null;
\EndIf
\State Initialization: $\ell=+\infty, \hbar=\lfloor (n-k)/2\rfloor$;
\For{ $\imath\in [1,n-2\hbar-1]$}
\State $B_{\imath}=[\imath,\imath+\hbar-1]$;
\State$//$\verb"Do Gaussian elimination to" $\mathcal{M}^{(\hbar+1)}_{B_{\imath}}$.
\State  $\widetilde{\mathcal{M}}^{(\hbar+1)}_{B_{\imath}}=$ RowsGaussianElimination$(\mathcal{M}^{(\hbar+1)}_{B_{\imath}})$;
\If{rank$(\widetilde{\mathcal{M}}^{(\hbar+1)}_{B_{\imath}})<\hbar+1$}
\State $  {\boxplus}_{\mathcal{M}^{(\hbar+1)}_{B_\imath}}=\{(e_1,f_1),\cdots,(e_v,f_v)\}$;
\State $\boxtimes_{\mathcal{M}^{(\hbar+1)}_{B_\imath}}=\{\lambda_1A_1+\lambda_2A_2|\forall \lambda_1,\lambda_2\in \mathbb{F}_{q^m},$
\State \hspace{12mm}$(\lambda_1,\lambda_2)\ne(0,0), \forall A_1,A_2 \in \boxplus_{\mathcal{M}^{(\hbar+1)}_{B_\imath}}\}$;
\State $ \widehat{\boxtimes}_{\mathcal{M}^{(\hbar+1)}_{B_\imath}}=\{(e,f)|(e,f)\in \boxtimes_{\mathcal{M}^{(\hbar+1)}_{B_\imath}},$
\State  \hspace{35mm} $\mathbf{G}e^T\ne\mathbf{G}f^T\}$;
\For{$\forall(e,f)\in \widehat{\boxtimes}_{\mathcal{M}^{(\hbar+1)}_{B_\imath}}$}
\If{$\max\{\textrm{bl}([e]),\textrm{bl}([f])\}<\ell$}
\State $\ell = \max\{\textrm{bl}([e]),\textrm{bl}([f])\}$;
\EndIf
\EndFor
\EndIf
\EndFor
\State return $\ell-1$;
\end{algorithmic}
\end{algorithm}

  \begin{table}
\caption{The true burst error correction ability of quantum RS codes.  The powers of the self-dual basis of $\mathbb{F}_{2^m}$ over $\mathbb{F}_2$ is denoted by
$\mathrm{SDB}_{m}$.}
\label{Computer Searching for QBECCRSs}
\centering
\begin{tabular}{|c|c|c|l|c|c|c|c|c| }
\hline
 $m$&  $n$&  $\mathrm{SDB}_{m}$ & $k$& $\mathcal{L}$ &\tabincell{c}{ $\ell_L$ in  \\ \cite{qian2015constructions,lin2004error}}   &  \tabincell{c}{  QRB in  \\ Theorem \ref{quantumReigerbound} }    \\
 \hline
 \multirow{2}{*}{$4$} &   \multirow{2}{*}{$15$}&\multirow{2}{*}{\{$6,9,11,14$\}} &   $5$& $8$ & $5$ & $10$  \\
  \cline{4-7}
  &   & &   $1$& $12$ & $9$ & $14$   \\
 \hline
 \multirow{9}{*}{$5$} &   \multirow{9}{*}{$31$}& \multirow{9}{*}{\{$3,5,11,22,24$\}}&   $23$& $7$ & $6$ & $10$  \\
 \cline{4-7}
  &   & &   $21$& $9$ & $6$ & $12$   \\
 \cline{4-7}
  &   & &   $19$& $12$ & $11$ & $15$   \\
 \cline{4-7}
  &   & &   $17$& $15$ & $11$ & $17$   \\
 \cline{4-7}
  &   & &   $15$& $17$ & $16$ & $20$   \\
 \cline{4-7}
 &   & &   $13$& $18$ & $16$ & $22$   \\
 \cline{4-7}
 &   & &   $11$& $22$ & $21$ & $25$   \\
 \cline{4-7}
&   &  &  $9$& $25$ & $21$ & $27$   \\
\cline{4-7}
&   &  &  $7$& $27$ & $26$ & $30$   \\
 \cline{4-7}
&   &  &  $5$& $29$ & $26$ & $32$   \\
 \cline{4-7}
&   &  &  $3$& $32$ & $31$ & $35$   \\
 \cline{4-7}
&   &   & $1$& $35$ & $31$ & $37$   \\
\hline
\multirow{24}{*}{$6$} &   \multirow{24}{*}{$63$}&   \multirow{24}{*}{\tabincell{c}{\{$13,23,26,44,$\\ $47,55$\}}}&  $55$& $8$ & $7$ & $12$  \\
  \cline{4-7}
   &   & &   $53$& $11$ & $7$ & $15$   \\
 \cline{4-7}
  &  &  &   $51$& $14$ & $13$ &  $18$   \\
 \cline{4-7}
  &   & &   $49$& $17$ & $13$ & $21$   \\
 \cline{4-7}
 &    & &   $47$& $20$ & $19$ & $24$   \\
 \cline{4-7}
 &   &  &  $45$& $23$ & $19$ & $27$   \\
 \cline{4-7}
&   &  &  $43$& $27$ & $25$ & $30$   \\
 \cline{4-7}
&   &  &  $41$& $29$ & $25$ & $33$   \\
 \cline{4-7}
&   &  &  $39$& $32$ & $31$ & $36$   \\
 \cline{4-7}
&   &  &  $37$& $33$ & $31$ & $39$   \\
\cline{4-7}
&   &  &  $35$& $38$ & $37$ & $42$   \\
 \cline{4-7}
&   &  &  $33$& $40$ & $37$ & $45$   \\
 \cline{4-7}
&   &   & $29$& $47$ & $43$ & $51$   \\
 \cline{4-7}
&   &   & $25$& $53$ & $49$ & $57$   \\
 \cline{4-7}
&   &   & $23$& $56$ & $55$ & $60$   \\
 \cline{4-7}
&   &  &  $21$& $59$ & $55$ & $63$   \\
 \cline{4-7}
&   &  &  $19$& $62$ & $61$ & $66$   \\
 \cline{4-7}
&   &  &  $17$& $66$ & $61$ & $69$   \\
 \cline{4-7}
&   &   & $15$& $68$ & $67$ & $72$   \\
 \cline{4-7}
&   &  &  $13$& $71$ & $67$ & $75$   \\
 \cline{4-7}
&   &   & $11$& $74$ & $73$ & $78$   \\
 \cline{4-7}
&   &  &  $9$& $76$ & $73$ & $81$   \\
 \cline{4-7}
&   &  &  $7$& $81$ & $79$ & $84$   \\
 \cline{4-7}
&   &  &  $5$& $84$ & $79$ & $87$   \\
 \cline{4-7}
&   &   & $3$& $86$ & $85$ & $90$   \\
 \cline{4-7}
&   &   & $1$& $89$ & $85$ & $93$   \\
\hline
\end{tabular}
\end{table}

\section{Error-Trapping Decoder of Quantum Cyclic Codes}
\label{quantumerrortrappingsection}
In this section, we present the quantum error-trapping decoder (QETD) for QCCs, and  we use the CSS-type QCCs to illustrate the proceess of decoding.  Let ${C}=[n,k]$ be a classical cyclic code over $\mathbb{F}_{2}$ such that ${C}^{\bot} \subseteq {C}$. Denote  the generator polynomial of ${C}$ by $\mathbf{g}(x)$, and denote   the parity-check matrix of ${C}$ by $\mathbf{H}$.    Let $\mathcal{Q}=[[n,\mathcal{K}=2k-n]]$ be a QCC constructed from ${C} $ by using the CSS construction.
Suppose that $\mathcal{Q}$ can correct any quantum burst error of length $\mathcal{L}$ or less. According to the quantum Reiger bound in Theorem \ref{quantumReigerbound}, we have $ \mathcal{L}\leq (n-\mathcal{K})/4$. In this section we present the quantum error-trapping decoder for QCCs to decode any quantum burst error of length up to $\mathcal{L}$. Similar to the classical error-trapping decoder, we show that QETD  can also correct additional quantum burst errors of length  $\mathcal{L}<l\leq (n-\mathcal{K})/2$. Such burst errors belong to the coset leaders of ${C}  $. In addition, we will show that QETD can also decode degenerate burst errors that belong to the coset of ${C}^{\bot}$.

  Let $|\psi\rangle$ be the encoded quantum state by $\mathcal{Q}$. Suppose that an $\mathbf{e}=(e_0,e_1,\cdots,e_{n-1})$  error pattern  is imposed on $|\psi\rangle$ during the transmission.  We define  $\mathbf{e}(x) \equiv e_0+e_1x+\cdots+e_{n-1}x^{n-1}$ as the error polynomial of $\mathbf{e}$. We perform   the syndrome measurement operation  by using the  stabilizer generators $\mathbf{H}$. Denote  the   syndrome information  by \begin{equation}
  \mathbf{S}\equiv\mathbf{H}\mathbf{e}^T=(s_0,s_1,\cdots,s_{r-1})^T
  \end{equation}   and denote by $\mathbf{S}(x)=s_0+s_1x+\cdots+s_{r-1}x^{r-1}$, where $r=n-k$. Moreover, the syndrome polynomial $\mathbf{S}(x)$ is equal to the remainder of dividing $\mathbf{e}(x)$ by the generator $\mathbf{g}(x)$, i.e., \begin{equation}\label{exgxsx}
\mathbf{e}(x)=u(x)\mathbf{g}(x)+\mathbf{S}(x).
\end{equation}

Let $\mathbf{e}=(e_0,e_1,\cdots,e_{n-1})$ be a correctable burst error of length $2\leq\ell\leq (n-\mathcal{K})/2=n-k$.
It is   natural to suppose that the burst length of $\mathbf{e}$ is less than or equal to $n-k$. If the quantum burst error $\mathbf{e}$ is  confined to the $n-k$ low-order positions, then $\mathbf{e}(x)=e_0+e_1x +\cdots+e_{n-k-1}x^{n-k-1}$.
 According to Eq. (\ref{exgxsx}), we know that $\mathbf{e}(x)=u(x)\mathbf{g}(x)+\mathbf{S}(x)$. Since the degree of $\mathbf{e}(x)$ is less than $n-k$, we have $\mathbf{S}(x)= \mathbf{e}(x)=e_0+e_1x +\cdots+e_{n-k-1}x^{n-k-1}$.  That is to say we can derive the burst error $\mathbf{e}(x)$ from the syndrome directly as long as $\mathbf{e}$ is  confined to the $n-k$ low-order positions. Moreover, all the burst errors that confined to the $n-k$ low-order positions range over all the   possible  $q^{n-k}$ error syndromes for ${C}$. This possibility is the key for the   decoding of QCCs.

\begin{figure}[!t]
\centering
\includegraphics[width=3.7in]{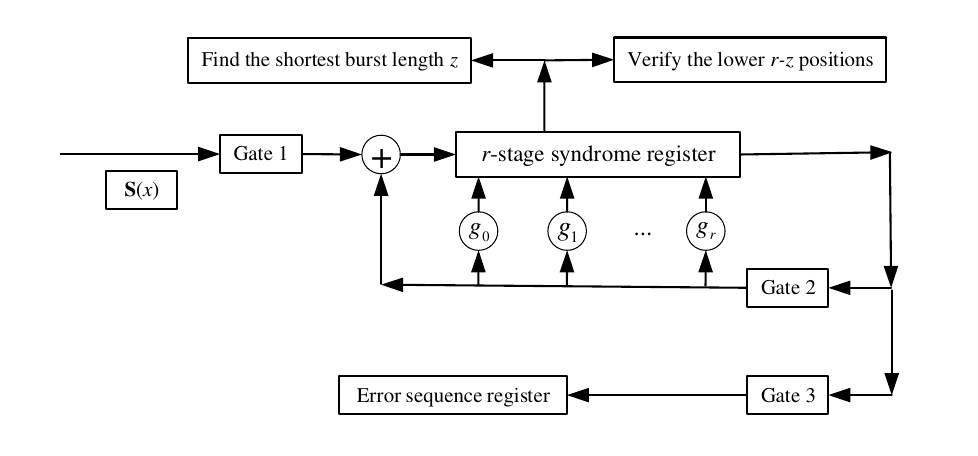}
\caption{Quantum error-trapping decoder for QCCs.}
\label{QETDCircuit}
\end{figure}

If  the   burst error $\mathbf{e}$ is not confined to the $n-k$ low-order positions, we    cyclically shift the syndrome  $\mathbf{S}$ by  a certain number of times   to  capture $\mathbf{e}$.
 If $0\leq \textbf{bl}(\mathbf{e})\leq \ell$, then $\mathbf{e}$ or an element in the coset   $\mathbf{e}+{C}^{\bot}$ can be captured in the $n-k$ low-order positions.
  If $\ell+1\leq \textbf{bl}(\mathbf{e})\leq n-k$, then  $\mathbf{e}$ may exceed  the burst error correction ability of $\mathcal{Q}$. However, we can still correct such burst error if it is a coset leader of $C$ or it   belongs to the coset of ${C}^{\bot}$. Before doing so, we first  cyclically shift the syndrome $n$ times to determine the shortest     burst   that is confined to the $n - k $ low-order positions   as in \cite{lin2004error,Gallager1968information}. Then we give a circuit for   QETD in  Fig. \ref{QETDCircuit} and present the whole process of QETD as follows.
\begin{itemize}
\item[(1)] The syndrome   $\textbf{S}(x)$ is shifted to the syndrome register with Gate 1   on and Gate 2 off.
\item[(2)] Shift the syndrome register with Gate 2 on and   a clock starts to count simultaneously. We use a counter $s$ to record the length $t$ of continuous zeros in the leftmost stages of the syndrome register, and $s$ is set to $0$ before the clock. If the length $t$ of continuous zeros  in the leftmost stages of the syndrome register  is larger than the previous one,       $s$ is updated by the length $t$ in the current clock.   After $n$ clocks, the counter $s$ is the longest  length of continuous zeros in the leftmost stages of the syndrome register. Then $z\equiv r-s$ is the shortest length of burst errors that appear in the $z$ rightmost stages of the syndrome register.

\item[(3)] Shift the syndrome register with Gate 2 on. As soon as the $r-z$   leftmost stages of the syndrome register contain all   zeros after the $i$th shift for $0\leq i\leq n-1$, the burst error is confined in the $z$ rightmost stages. Then  Gate 2 is   turned off.
 \item[(4)] The syndrome register is   shifted by $z$ times with Gate 3  on.    Then the error burst  is confined  to the error sequence register and Gate 3 is turned off.  We need to continue to shift the  error register so that the error burst is put in the right positions. With Gate 3 on, the error sequence register is cyclically  shifted by $[(n-z-i) \mod n]$ times.

    \end{itemize}

  The decoding circuit in Fig. \ref{QETDCircuit} is easy to be implemented by using the linear shift register.  In steps (2) and (3), the shift registers both run in   $O(n)$ time. In step (4), the syndrome  register also runs in $O(n)$ time. Therefore the total time complexity of the QETD algorithm is linear.
 On the other hand, for the purpose of numerical simulations, we give a simplified QETD algorithm for QCCs in Algorithm \ref{ErTrDeQuan}.  It should be noted that  the time complexity of Algorithm \ref{ErTrDeQuan} is indeed $O(n^2)$ which is   higher than the complexity of the circuit level decoder in Fig. \ref{QETDCircuit}. But Algorithm 3 is easier  to be simulated in computers by using   high level programming languages such as C/C++, Magma and Python, etc.

 \renewcommand{\algorithmicrequire}{\textbf{Input:}}
\renewcommand{\algorithmicensure}{\textbf{Output:}}
\begin{algorithm}[H]
\caption{Quantum Error-trapping Decoder    for  Quantum Cyclic Codes.}
\label{ErTrDeQuan}
\begin{algorithmic}[1]
\Require
  $\textbf{S}(x), \textbf{g}(x)$;
\Ensure The decoded error sequence ${e}_X$.
\State            Initialization: $\textbf{Z}  = +\infty$, $   {v}=0$;
\State $//$ \verb"Calculate the shortest burst that is" \verb"confined in the right most stages of" \verb"the syndrome".
\For{  $i\in [0,n-1]$}
\State $\textbf{S}^{(i)}(x)=x^{i}\textbf{S}(x) \mod  \textbf{g}(x)$
\If  { $\textbf{S}^{(i)}(n-k) == 1 \&\& \textbf{Z} >\textbf{bl}(\textbf{S}^{(i)}(x))$}
 \State  $\textbf{Z} =\textbf{bl}(\textbf{S}^{(i)}(x));$
 \State $v=i;$
\EndIf
\EndFor
\State $//$ \verb"Shift the error sequence to the right" \verb"position."
 \State  $\textbf{e}(x)=x^{-v}(x^v\textbf{S}(x) \mod \textbf{g}(x))\mod (x^n-1);$ \\
\Return $\textbf{e}(x)$;
\end{algorithmic}
\end{algorithm}

The QETD in Algorithm   \ref{ErTrDeQuan} can decode all   correctable burst errors of length $\ell \leq   \mathcal{L}$ which indicates the burst error correction ability of the QCCs. Moreover, the QETD  algorithm can also decode all correctable burst errors of length up to $(n-\mathcal{K})/2$ that are the coset leaders, which have the shortest burst length in the coset. Therefore, the QETD  algorithm  is quantum maximum likelihood decoding of nondegenerate burst errors according to \cite{hsieh2011np,lin2004error,Gallager1968information}. Furthermore, the QETD can also decode     degenerate errors belong to the coset of $C^\bot$. In addition, QETD   in Algorithm   \ref{ErTrDeQuan}  is also available for Hermitian-ype QCCs. In Table \ref{degenerateQETD}, we list the numerical results of several QCCs for correcting quantum burst errors. Denote $\mathcal{N}_D$, $\mathcal{N}_0$, and $\mathcal{N}$ by the numbers of all correctable burst errors,  all correctable nondegenerate burst errors and total errors of length $n\leq  (n-\mathcal{K})/2$, respectively. We exhaustively traverse all the burst errors of length $\ell \le (n-\mathcal{K})/2$ and count the numbers of burst errors that are successfully decoded by QETD.  It is shown in Table \ref{degenerateQETD} that  QETD can correct   more degenerate  burst errors than nondegenerate ones. As the code length grows, the ratio $\mathcal{N}_D/\mathcal{N}_0$ becomes extremely larger and thus QETD can correct much more degenerate burst errors than nondegenerate ones.
As an example,   nondegenerate burst errors that can be decoded by the   $Q=[[29,1]]$ QCC  are   only   $4.4\%$ of the total burst errors  of length $\ell\leq 12$.  However, $Q$ can decode  degenerate burst errors up to $26.9\%$ of the total burst errors of length $\ell\leq 12$.

\begin{table*}
\caption{Numercial Results of Quantum Cyclic Codes for Decoding Burst Errors by Using the Quantum Error-Trapping Decoder.}
\label{degenerateQETD}
\centering
\begin{tabular}[c]{|l|l|l|l|l|l|l|l| }
\hline
 $[[n,k]]$& $\mathcal{N}_D$   &  $\mathcal{N}_0$ &  $\mathcal{N}$ &  $\mathcal{N}_D/\mathcal{N}$&  $\mathcal{N}_0/\mathcal{N}$&$\mathcal{N}_D/\mathcal{N}_0$&  Generator Polynomials  $g$ \\
 \hline
 $[[5,1]]$&$15$&$15$ &$51$& $29.4\%$&$29.4\%$&$1$
 & $g=(\textbf{1}^{2}\textbf{2}^{1}\textbf{1}^0)$\\
\hline
 $[[7,1]]$&$72$&$57$& $255$&$28.2\%$&$22.3\%$&$1.26$&$g=(\textbf{1}^{3}\textbf{1}^{1}\textbf{1}^0)$ \\
\hline
$[[13,1]]$&$7623$&$2865$ &$25599$&$29.7\%$&$11.1\%$&$2.66$&$g=(\textbf{1}^{6}\textbf{2}^{5}\textbf{3}^{3}\textbf{2}^{1}\textbf{1}^0)$ \\
\hline
$[[17,1]]$&$1.45401\rm{E}{5}$&$4.1064\rm{E}{4}$ &$5.07903\rm{E}{5}$&$28.6\%$&$8\%$&$3.54$&$g=(\textbf{1}^{8}\textbf{3}^{7}\textbf{3}^{5}\textbf{3}^{4}\textbf{3}^{3}\textbf{3}^{1}\textbf{1}^0)$ \\
\hline
$[[23,1]]$&$1.1514471\rm{E}{7}$&$2.395308\rm{E}{6}$ &$4.1943039\rm{E}{7}$&$27.4\%$&$5.7\%$&$4.81$&$g=(\textbf{1}^{11}\textbf{1}^{9}\textbf{1}^{7}\textbf{1}^6\textbf{1}^5\textbf{1}^1\textbf{1}^0)$ \\
\hline
$[[25,1]]$&$4.9269693\rm{E}{7}$&$9.363588\rm{E}{6}$&$1.80355071\rm{E}{8}$ &$27.3\%$&$5.1\%$&$5.26$&$g=(\textbf{1}^{12}\textbf{2}^{11}\textbf{1}^{10}\textbf{2}^{7}\textbf{3}^{6}\textbf{2}^{5} \textbf{1}^{2}\textbf{2}^{1}\textbf{1}^0)$ \\
\hline
$[[29,1]]$&$8.86214133\rm{E}{8}$ &$1.44826293\rm{E}{8}$&$3.288334248\rm{E}{9}$ &$26.9\%$&$4.4\%$&$6.12$&$g=(\textbf{1}^{14}\textbf{2}^{13}\textbf{2}^{11}\textbf{3}^{10}\textbf{1}^{9}\textbf{3}^{8}\textbf{2}^{7}
\textbf{3}^{6}\textbf{1}^{5}\textbf{3}^{4} \textbf{2}^{3}\textbf{2}^{1}\textbf{1}^0)$ \\
\hline
\end{tabular}
\end{table*}
\section{Conclusion and Discussion}
\label{conclusionanddiscussion}
In this paper, we characterized the issue of burst error correction   of quantum cyclic codes.   We proposed a polynomial-time algorithm  to determine the burst error correction limit of general QCCs, and then we derived many optimal or nearly optimal QCCs.  We   proposed a polynomial-time algorithm to determine the true burst error correction limit of quantum RS codes. We showed that   quantum RS codes can beat the previous lower bound for burst error correction.  At last, we proposed quantum error-trapping decoder  for  correcting burst errors of QCCs.  We showed that the quantum error-trapping decoder can not only decode all the burst errors that are coset leaders but also can decode   degenerate errors belong to the coset of coset leaders. The numerical results showed that QETD can decode much more degenerate burst errors than nondegenerate ones.

 Regarding  the future work, how to determine the burst error correction ability of QCCs with mathematical methods is quite useful. Although the time complexity of Algorithms \ref{BurstLengthOfQCCs}$\&$\ref{BurstLOfQRS}  is polynomial, the   exhaustive searching complexity is high   when the code length is relatively large. On the other hand, whether the quantum error-trapping decoder is degenerate, quantum maximum likelihood decoding is unknown. It is an interesting problem to find degenerate quantum maximum likelihood decoding for quantum cyclic codes.

\section*{Acknowledgment}
The authors would like to thank the Editor and the anonymous referees for their valuable comments and suggestions       for improving the presentation of their article.


\ifCLASSOPTIONcaptionsoff
  \newpage
\fi



%
\bibliographystyle{IEEEtran}
\bibliography{IEEEabrv,thebibfile}
%







\appendices
\section{Proof of the Quantum Reiger Bound}
\label{appendixA}

 \begin{lemmas}[No-Cloning Bound]
  \label{no-cloning bound for QBECCs}
  For an arbitrary $\ell$ burst error correction code $C=[[n,k\geq 1]]$   exists only
  if
  \begin{equation}
  n>4\ell.
  \end{equation}
  \end{lemmas}
\begin{IEEEproof}
Suppose that there exists a code  $C'=[[n,k\geq 1]]$ with $2\leq n\leq4\ell$. After encoding $k$ qubits into $n$ ones, we split the
encoded block into two sub-blocks, one contains the first $\lfloor \frac{n}{2}\rfloor$ qubits and the other contains the rest of the
$n-\lfloor \frac{n}{2}\rfloor$ qubits.

If we append $\lfloor \frac{n}{2}\rfloor$  ancilla qubits $|0\cdots0\rangle$  to the first sub-block, and
append $n-\lfloor \frac{n}{2}\rfloor$  ancilla qubits $|0\cdots0\rangle$  to the second sub-block, then the original encoded
block has spawned two offspring, the first one with  located burst errors of length
at most $\lfloor \frac{n}{2}\rfloor$, and the second one with located burst errors of length
at most  $n-\lfloor \frac{n}{2}\rfloor$. If we were able to
correct the two located burst errors in each of the offspring (see Lemma \ref{Located errors for QBECCs}), we would obtain two
identical copies of the parent encoded block, which is a contradiction with the quantum no-cloning theorem \cite{nielsen2010quantum}. Therefore we must have $n>4\ell.$
\end{IEEEproof}

\begin{lemmas}[Located Burst Errors]
\label{Located errors for QBECCs}
For a  QECC $Q=[[n,k]]$ that corrects arbitrary burst errors of length  $\ell$ or less can correct  located burst errors of length at most $2\ell$.
\end{lemmas}

    \begin{IEEEproof}
  Denote   an arbitrary error of length $n$ by
  \begin{equation}
  e= e_1\otimes\ldots\otimes e_x\otimes\ldots\otimes e_y\otimes\ldots\otimes e_n ,
  \end{equation}
where $1\leq x<y\leq n, y-x+1=2\ell$, and $e_i(1\leq i\leq n)$ are Pauli matrices.
The set $E(x,y)$ of burst errors to be corrected is the set of all Pauli operators, {where each acts trivially on the qubits $1$ to $x-1$ and on the qubits $y+1$ to $n$ (except $x=1$ and $y=n$)}. Then each error  in $E(x,y)$ has a burst length of at most $2\ell$.
But now, for each $E_a$ and $E_b$ in $E(x,y)$, the product $E_a^\dagger E_b$
also has a burst of length at most $2\ell$. Therefore, the burst error-correcting criterion (\ref{burst-error-correction criterion})
is satisfied for all $E_{a,b}\in E$, provided $Q$ is an $\ell$ burst error correction code.
  \end{IEEEproof}

\begin{IEEEproof}[Proof of Theorem \ref{quantumReigerbound}]
The proof follows closely by that of the quantum Singleton bound given by Preskill (see \cite[p.32]{preskill1998physics} and \cite[p.568]{nielsen2010quantum}).

First of all, Lemma~\ref{no-cloning bound for QBECCs}  says that if $Q$ can correct $\ell$ burst errors, then it must satisfy $n>4\ell$, a consequence following from the quantum no-cloning principle.

Then we introduce a $k$-qubit ancilla system $A$, and construct a pure state $|\Psi\rangle_{AQ}$ that is maximally entangled between the system $A$ and the $2^k$ codewords of the $[[n,k]]$ QBECC $Q$:
\begin{equation}
|\Psi\rangle_{AQ}=\frac{1}{\sqrt{2^k}}\sum|x\rangle_A|{x}\rangle_Q,
\end{equation}
where $\{|x\rangle_A\}$ denotes an orthonormal basis for the $2^k$-dimensional Hilbert space of the ancilla, and $\{|x\rangle_Q\}$ denotes an orthonormal basis for the $2^k$-dimensional code subspace. 
It is obvious that
\begin{equation}
S(A)_\Psi=k=S(Q)_\Psi,
\end{equation}
where $S(A)_\rho= -\text{Tr} \rho_A \log \rho_A $ is the von Neumann entropy of a density operator $\rho_A$.


Next we divide the $n$-qubit QBECC $Q $ into three disjoint parts so that  $Q^{(1)}$ and $Q^{(2)}$ consist of $2\ell$ qubits each and $Q^{(3)}$ consists of the remaining $n-4\ell$ qubits. If we trace out $Q^{(2)}$ and $Q^{(3)}$, the reduced density matrix that we obtained must contain no correlations between $Q^{(1)}$ and the ancilla $A$, a consequence following from Lemma \ref{Located errors for QBECCs} in the Appendix. This means that the entropy of
system $AQ^{(1)}$ is additive:
\begin{equation}
S( Q^{(2)}Q^{(3)})_\Psi=S(AQ^{(1)})_\Psi=S(A)_\Psi+S(Q^{(1)})_\Psi.
\end{equation}
Similarly,
\begin{equation}
S( Q^{(1)}Q^{(3)})_\Psi=S(AQ^{(2)})_\Psi=S(A)_\Psi+S(Q^{(2)})_\Psi.
\end{equation}
Furthermore, in general, the von Neumann entropy is subadditive, so that
\begin{eqnarray}
S( Q^{(1)}Q^{(3)})_\Psi\leq S(Q^{(1)})_\Psi +S(Q^{(3)})_\Psi\\
S( Q^{(2)}Q^{(3)})_\Psi\leq S(Q^{(2)})_\Psi +S(Q^{(3)})_\Psi.
\end{eqnarray}
Combining these inequalities with the equalities above, we find
\begin{eqnarray}
S(A)+S(Q^{(2)})_\Psi\leq S(Q^{(1)})_\Psi +S(Q^{(3)})_\Psi\\
S(A)+S(Q^{(1)})_\Psi\leq S(Q^{(2)})_\Psi +S(Q^{(3)})_\Psi.
\end{eqnarray}
Both inequalities can be simultaneously satisfied only if
\begin{equation}
S(A)_\Psi\leq S(Q^{(3)})_\Psi.
\end{equation}
Finally, we have
\begin{equation}
S(A)_\Psi=k\leq n-4\ell,
\end{equation}
since $S(Q^{(3)})$ is bounded above by its dimension $n - 4\ell$. We then conclude the quantum Reiger bound.
\end{IEEEproof}
\section{Proof of Theorem \ref{theoremquantumburst1}}
\label{appendixtheoremquantumburst1}

\begin{IEEEproof}[Proof of Theorem \ref{theoremquantumburst1}]
If item 1) holds,  we know that ${C}$ can correct any burst error of length $\ell$ or less according to Lemma \ref{classicalburstcyclic}. If   $\mathcal{M}^{(\ell)}_{A_{\imath}}$   is not of full rank for some $1\leq \imath\leq n-2\ell+1$, it means that  a number of  columns in  $\mathcal{M}^{(\ell)}_{A_{\imath}}$    are linearly dependent. It is known that every set of linearly dependent columns in $\mathcal{M}^{(\ell)}_{A_{\imath}}$ corresponds to a pair of burst errors $(e,f)$ such that $He^T=Hf^T$. We define a set of all such burst errors as $ \mathcal{E}_{\mathcal{M}^{(\ell)}_{A_{\imath}}}$.
   We need to verify that whether such burst errors in $ \mathcal{E}_{\mathcal{M}^{(\ell)}_{A_{\imath}}}$  are degenerate or not.

For   each   $\beta_i=a_{i1}\alpha_1+\cdots+a_{iu}\alpha_v$ ($1\leq i\leq v$) in $\widehat{\mathscr{D}}_H$ and the corresponding burst error $(e_i,f_i)\in\boxplus_{\mathcal{M}^{(\ell)}_{A_{\imath}}}$, we have $\mathbf{H}e_i^T=\mathbf{H}f_i^T$. If $\mathbf{G}^\dagger e_i^T=\mathbf{G}^\dagger f_i^T$, then $e_i$ and $f_i$ are degenerate. Let $2\leq w\leq v$ and let $1\leq i_1< \cdots<i_{w}\leq v$.
If  $\beta_{i_1},\cdots,  \beta_{i_{w-1}}$,  and $\beta_{i_w}$  are linearly dependent, then we have $\beta_{i_1}=b_{i_1}\beta_{i_2}+\cdots+b_{i_w}\beta_{i_w}$, where $b_{i_j} \in  \mathbb{F}_{q^2}$ for $1\leq j\leq w$. Denote  the burst error that corresponds to the positions of columns $\beta_{i_1},\cdots,  \beta_{i_{w-1}}$ in $\mathbf{H}$ by $e_L$. Then we have $e_L=e_{i_1}+\cdots+e_{i_w}$ and $\mathbf{H}e_L^T= \mathbf{H} e_{i_1}^T+\cdots+ \mathbf{H} e_{i_w}^T=\mathbf{H}f_{i_1}^T+\cdots+ \mathbf{H} f_{i_w}^T= \mathbf{H} (f_{i_1}^T+\cdots+ f_{i_w}^T)$. Denote by $f_L=f_{i_1}+\cdots+f_{i_w}$.  Then we   have $\mathbf{G}^\dagger e_L^T=\mathbf{G}^\dagger e_{i_1}^T+\cdots+\mathbf{G}^\dagger e_{i_w}^T=\mathbf{G}^\dagger f_{i_1}^T+\cdots+\mathbf{G}^\dagger f_{i_w}^T=\mathbf{G}^\dagger f_L^T$. Thus $e_L$ and $f_L$ are degenerate errors. Therefore, if $\mathcal{M}^{(\ell)}_{A_{\imath}}$   is not of full rank and
$\mathbf{G}^\dagger e^T=\mathbf{G}^\dagger f^T$ for all $(e,f)\in\boxplus_{\mathcal{M}^{(\ell)}_{A_{\imath}}}$, then all the errors in $ \mathcal{E}_{\mathcal{M}^{(\ell)}_{A_{\imath}}}$ are degenerate.

According to Lemma \ref{Hermitian QBECCs}, we can construct   a ${Q}=[[n,2k-n]]$ QBECC which can correct any quantum burst error of length $\mathcal{L}$ or less.
  
 \end{IEEEproof}
\section{Proofs of Lemma \ref{relationshipofRSandExpan}, Lemma \ref{RSMBl}, and Theorem \ref{theoremQRSburstlimit}}
\label{appendixrelationshipofRSandExpan}
 \begin{IEEEproof}[Proof of Lemma \ref{relationshipofRSandExpan}]
 For each $\mu=(\mu_1,\cdots,\mu_n)\in C$, there is
\begin{equation}
\mu_{i}=\sum_{j=1}^m u_{ij}\alpha_j,1\leq i\leq n.
 \end{equation}
Let $[\mu_{i}]=(u_{i1},\cdots,u_{im})$ for $1\leq i\leq n$. Then there is $[\mu]=([\mu_{1}],\cdots,[\mu_{n}])\in\mathbf{D}$. For arbitrary two codewords $\mu^{(1)}=(\mu_1^{(1)},\cdots,\mu_n^{(1)}) \in C$ and $\mu^{(2)}=(\mu_1^{(2)},\cdots,\mu_n^{(2)})\in C$, if $\mu^{(1)}\ne\mu^{(2)}$,  we   have
$\mu_\mathbf{i}^{(1)} \ne \mu_\mathbf{i}^{(2)}$ for some $1\leq \mathbf{i}\leq n$. Then there must be $[\mu^{(1)}] \ne [\mu^{(2)}]$.

Let $\mathbf{v}^{(1)}=(\mathbf{v}_1^{(1)},\cdots,\mathbf{v}_n^{(1)})$ and   $\mathbf{v}^{(2)}=(\mathbf{v}_1^{(2)},\cdots,\mathbf{v}_n^{(2)})  $ be arbitrary two codewords of $\mathbf{D}$, where $\mathbf{v}_i^{(1)}=(\mathbf{v}_{i1}^{(1)},\cdots,\mathbf{v}_{im}^{(1)})$  and $\mathbf{v}_i^{(2)}=(\mathbf{v}_{i1}^{(2)},\cdots,\mathbf{v}_{im}^{(2)})$ for $1\leq i\leq n$.  Then $\mathbf{v}^{(1)}$ and $\mathbf{v}^{(2)}$ correspond to two codewords
 $\nu^{(1)}=(\nu_1^{(1)},\cdots,\nu_n^{(1)})\in C$ and $\nu^{(2)}=(\nu_1^{(2)},\cdots,\nu_n^{(2)})\in C$, respectively, where $\nu_i^{(1)}=\sum_{j=1}^m \mathbf{v}_{ij}^{(1)} \alpha_j $ and $\nu_i^{(2)}=\sum_{j=1}^m \mathbf{v}_{ij}^{(2)} \alpha_j $ for $1\leq i\leq n$.  If $ \mathbf{v}^{(1)} \ne  \mathbf{v}^{(2)} $, there is
$\mathbf{v}_\mathbf{i}^{(1)} \ne \mathbf{v}_\mathbf{i}^{(2)}$ for some $1\leq \mathbf{i}\leq n$, and then there is
 $\mathbf{v}_{\mathbf{i}\mathbf{j}}^{(1)} \ne \mathbf{v}_{\mathbf{i}\mathbf{j}}^{(2)}$ for some $1\leq \mathbf{j}\leq m$. Suppose that    $\nu_\mathbf{i}^{(1)}=\nu_\mathbf{i}^{(2)}$, then there is
\begin{equation}
\sum_{j=1}^m(\nu_{\mathbf{i}j}^{(1)}-\nu_{\mathbf{i}j}^{(2)})\alpha_j= 0.
\end{equation}
Then there is $ \nu_{\mathbf{i}j}^{(1)}=\nu_{\mathbf{i}j}^{(2)} $ for all $1\leq j\leq m$ which is a contradiction with $ \nu_{\mathbf{i}\mathbf{j}}^{(1)}\ne\nu_{\mathbf{i}\mathbf{j}}^{(2)} $. Therefore we must have $\nu_\mathbf{i}^{(1)}\ne\nu_\mathbf{i}^{(2)}$ and then we have $\nu^{(1)}\ne\nu^{(2)}$.
 
 \end{IEEEproof}

\begin{IEEEproof}[Proof of Lemma \ref{RSMBl}]
Let $\hat{B}_{\imath}=[\imath,\imath+\hbar-1]$, where $1\leq \imath\leq n-2\hbar-1$. We have
 \begin{equation}
\label{MbhandMbh1}
\mathcal{M}^{(\hbar+1)}_{ {B}_\imath} =
\left(
\begin{array}{cc}
\mathcal{M}^{(\hbar+1)}_{\hat{B}_\imath}&A_{\imath+\hbar} \\
  \end{array}
\right)
\end{equation}
and
\begin{equation}
\label{MbhandMbh1}
\mathcal{M}^{ \hbar }_{\hat{B}_\imath} =
\left(
\begin{array}{cc}
\mathcal{M}^{(\hbar+1)}_{\hat{B}_\imath}  \\
B_{r-\hbar}
  \end{array}
\right),
\end{equation}
where $A_{\imath+\hbar}$ is the $(\imath+\hbar)$th  column  of $\mathcal{M}^{(\hbar+1)}_{ {B}_\imath}$, and $B_{r-\hbar}$ is the $(r-\hbar)$th  row of $\mathcal{M}^{ \hbar }_{\hat{B}_\imath}$. Since RS codes saturate the Reiger bound, $\mathcal{M}^{ \hbar }_{\hat{B}_\imath}$ is of full rank and is equal to $\hbar$. Then $\mathcal{M}^{(\hbar+1)}_{\hat{B}_\imath} $ must be greater than or equal to $\hbar-1$. Therefore we have $\hbar-1\leq rank(\mathcal{M}^{(\hbar+1)}_{B_\imath}) \leq \hbar$.
\end{IEEEproof}

\begin{IEEEproof}[Proof of   Theorem \ref{theoremQRSburstlimit}]
Let ${D}=[{C}]$ be the $q$-ary expansion of ${C}$ under the self-dual basis. Then there is ${D}^\bot\subseteq {D}$ according to Lemma \ref{dualbinaryexpansion}. Moreover, ${D}$ can correct any quantum burst error of length $(\hbar-1)m+1$ or less according to Lemma \ref{binaryexpanburstbound}.
Let $\widetilde{e}$ and $\widetilde{f}$ be arbitrary two burst error over $\mathbb{F}_q$ such that\begin{equation}
(\hbar-1)m+1\leq \textrm{bl}(\widetilde{e})\leq\mathcal{L}, \textrm{ and } 1\leq\textrm{bl}(\widetilde{f})\leq\mathcal{L}.
\end{equation}
We map $\widetilde{e}$ and $\widetilde{f}$ to two elements over finite field $\mathbb{F}_{q^m}$ and   denote them by $ e $ and $ f $, respectively, and then, we have $ \widetilde{e}=[e] $ and  $ \widetilde{f}=[f] $. We also have
\begin{equation}
 \hbar   \leq \textrm{bl}(e)\leq \hbar+1,  \textrm{ and } 1\leq\textrm{bl}(f)\leq \hbar +1.
\end{equation}

Suppose that $\widetilde{e}+\widetilde{f} \in {D}\setminus {D}^{\bot}$, then $ e + f  \in {C}\setminus {C}^{\bot}$ according to Lemma \ref{relationshipofRSandExpan}.  It is a contradiction with Eq. (\ref{RSImageburstMax}). Then there exists a $Q=[[n,2k-n]]$ quantum RS code which  can correct any quantum burst error of length $\mathcal{L}$ or less according to Lemma \ref{CSS QBECCs} and Lemma \ref{dualbinaryexpansion}.

\end{IEEEproof}

 \begin{IEEEbiographynophoto}{Jihao Fan}
received the B.S. degree in mathematics from Lanzhou University, Lanzhou, China, in 2009, and the Ph.D. degree in computer software and theory from Southeast University, Nanjing, China, in 2016. He was a joint-training Ph.D student at Sydney University, Australia from 2014 to 2015. He is currently an Associate Professor with the Nanjing University of Science and Technology, Nanjing, China. His research interests include classical and quantum coding theory, information theory, and code-based masking. \end{IEEEbiographynophoto}

\begin{IEEEbiographynophoto}{Min-Hsiu Hsieh}
 has been the director of the Hon Hai Quantum Computing Research Center in Taiwan since January 2021. Prior to this role, he was an Associate Professor at the University of Technology Sydney in Australia and a member of the UTS Centre for Quantum Software and Information from 2014 to 2020. From 2010 to 2012, he was a postdoctoral research fellow at the University of Cambridge in the UK, where he was affiliated with the Centre for Quantum Information and Foundations. Before that, he worked as a research scientist with the ERATO-SORST Quantum Computation and Information Project in Japan, which was funded by the Japan Science and Technology Agency. During this time, he was also affiliated with the Department of Computer Science at the University of Tokyo. From 2014 to 2018, he held an Australian Research Council Future Fellowship. His research interests include general topics in quantum information and computation.
\end{IEEEbiographynophoto}

\end{document}